\documentclass[aps,prb,superscriptaddress,floatfix,preprint,letterpaper]{revtex4-1}
\pdfoutput=1

\usepackage[justification=raggedright, margin=5mm,font={sf,small},labelfont=bf, format=plain]{caption}

\begin{document}

\begin{titlepage}

\title{
Topological order:
from long-range entangled quantum matter\\
to a unified origin of light and electrons
}

\author{Xiao-Gang Wen}
\affiliation{Perimeter Institute for Theoretical Physics, Waterloo, Ontario, N2L 2Y5 Canada}
\affiliation{Department of Physics, Massachusetts Institute of
Technology, Cambridge, Massachusetts 02139, USA}
\affiliation{Institute for Advanced Study, Tsinghua University,
Beijing, 100084, P. R. China}

\begin{abstract}
In primary school, we were told that there are four states of matter: solid,
liquid, gas, and plasma.  In college, we learned that there are much more than
four states of matter. For example, there are ferromagnetic states as revealed
by the phenomenon of magnetization and superfluid states as defined by the
phenomenon of zero-viscosity.  The various phases in our colorful world are so
rich that it is amazing that they can be understood systematically by the
symmetry breaking theory of Landau.  In this paper, we will review the progress
in last 20 -- 30 years, during which we discovered that there are even more
interesting phases that are beyond Landau symmetry breaking
theory.  We discuss new ``topological'' phenomena, such as topological
degeneracy, that reveal the existence of those new phases -- topologically
ordered phases.  Just like zero-viscosity defines the superfluid order, the new
``topological'' phenomena define the topological order at macroscopic level.
As a new type of order, topological order 
requires a new mathematical frame work, such as fusion
category and group cohomology, to describe it.
More recently, we find that, at microscopical level, topological order is due
to long-range quantum entanglements, just like fermion superfluid is due to
fermion-pair condensation.  
Long-range quantum entanglements lead to many amazing emergent phenomena, such
as fractional quantum numbers, fractional/non-Abelian statistics, and perfect
conducting boundary channels.  Long-range quantum entanglements
can even provide a unified origin of light and electrons (or more
generally, gauge interactions and Fermi statistics): 
%quantum gauge fields are not curvatures of fiber bundle but 
light waves (gauge fields) are fluctuations of long-range entanglements, and
electrons (fermions) 
%are not a Grassmann fields but 
are defects of long-range entanglements.  Long-range quantum entanglements
represent a new chapter and a future direction
of condensed matter physics, or even physics in general.

\end{abstract}

%\pacs{71.27.+a, 02.40.Re}

\maketitle

\end{titlepage}

{\small \setcounter{tocdepth}{1} \tableofcontents }

~

~

\noindent
\emph{Symmetry is beautiful and rich.\\
Quantum entanglement is even more beautiful and richer.}

\section{Introduction}

\subsection{Phases of matter and Landau symmetry breaking theory}

Although all matter is formed by only three kinds of particles: electrons,
protons and neutrons, matter can have many different properties and appear in
many different forms, such as solid, liquid, conductor, insulator, superfluid,
magnet, etc. According to  the principle of emergence in condensed matter
physics, the rich properties of materials originate from the rich ways in which
the particles are organized in the materials.  Those different organizations of
the particles are formally called the orders in the materials.

\begin{figure}[t]
\centerline{
\includegraphics[scale=0.5]{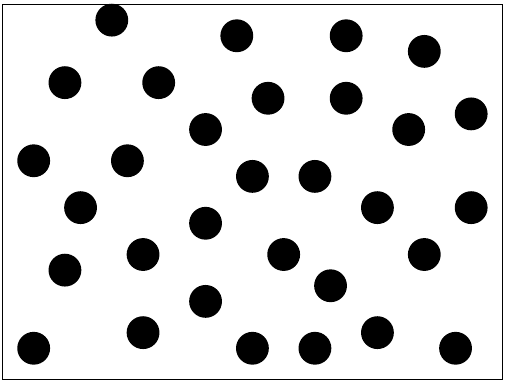}
\hfil
\includegraphics[scale=0.5]{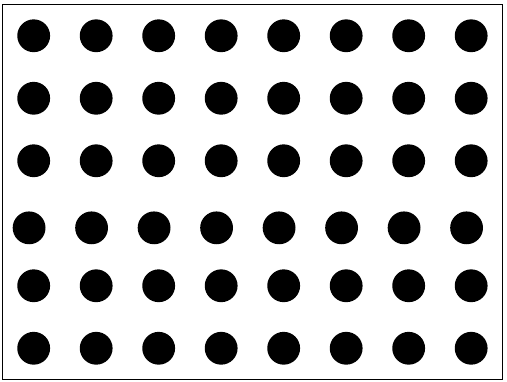}
}
\centerline{
\hfil
(a)
\hfil\hfil
(b)
\hfil
}
\caption{
(a) Particles in liquids do not have fixed relative positions.
They fluctuate freely and have a random but uniform distribution.
(b) Particles in solids form a fixed regular lattice.
}
\label{cryliq1}
\end{figure}

\begin{figure}[tb]
\centerline{
\includegraphics[scale=0.6]{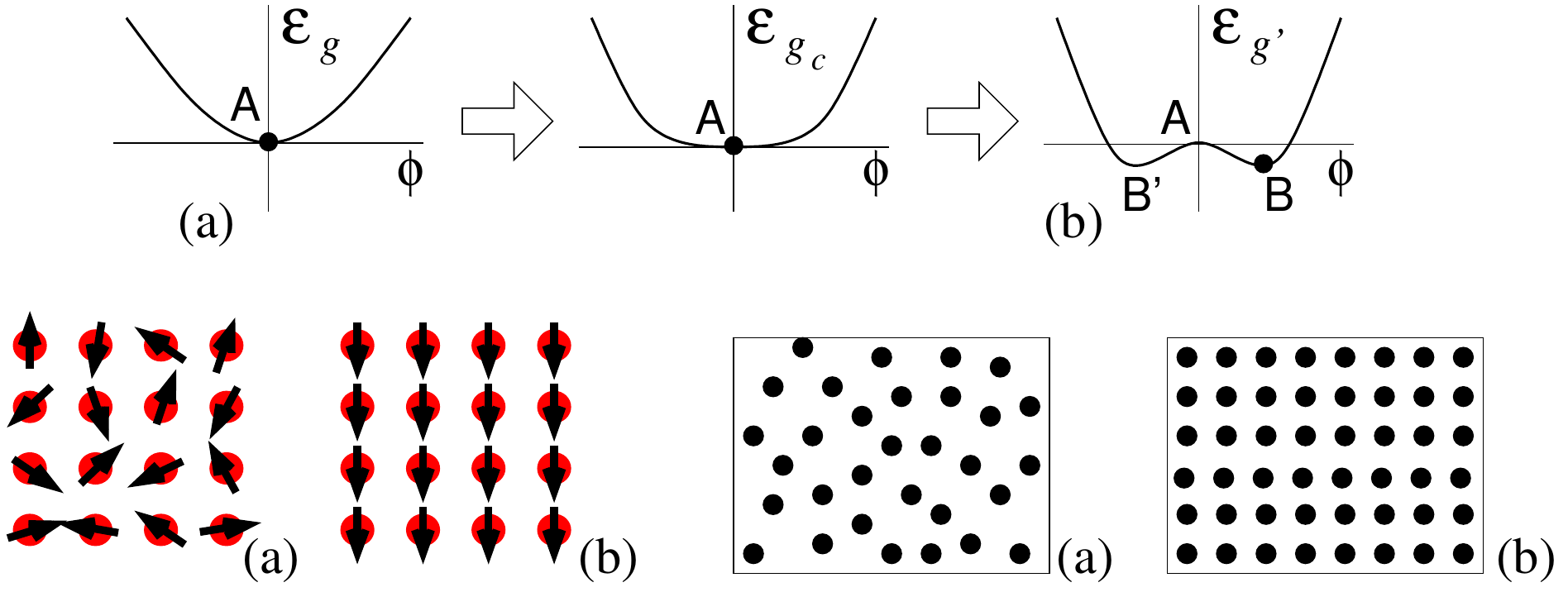}
}
\caption{ (a) Disordered states that do not break the symmetry.  (b) Ordered
states that spontaneously break the symmetry.  The energy function
$\veps_g(\phi)$ has a symmetry $\phi \to -\phi$:
$\veps_g(\phi)=\veps_g(-\phi)$.  However, as we change the parameter $g$, the
minimal energy state (the ground state) may respect the symmetry (a), or may
not respect the symmetry (b). This is the essence of spontaneous symmetry
breaking.  }
\label{symmbrk}
\end{figure}

For example, particles have a random distribution in a liquid (see Fig.
\ref{cryliq1}a), so a liquid remains the same as we displace it by an arbitrary
distance.  We say that a liquid has a ``continuous translation symmetry''.
After a phase transition, a liquid can turn into a crystal.  In a crystal,
particles organize into a regular array (a lattice) (see Fig. \ref{cryliq1}b).
A lattice remains unchanged only when we displace it by a particular distance
(integer times of lattice constant), so a crystal has only ``discrete
translation symmetry''.  The phase transition between a liquid and a crystal is
a transition that reduces the continuous translation symmetry of the liquid to
the discrete symmetry of the crystal. Such change in symmetry is called
``spontaneous symmetry breaking''.  We note that the equation of motions that
govern the dynamics of the particles respect the continuous translation
symmetry for both cases of liquid and crystal.  However, in the case of
crystal, the stronger interaction makes the particles to prefer being separated
by a fixed distance and a fixed angle.  This makes particles to break the
continuous translation symmetry down to discrete translation symmetry
``spontaneously'' in order to choose a low energy configuration (see Fig.
\ref{symmbrk}).  Therefore, the essence of the difference between liquids and
crystals is that the organizations of particles have different
\emph{symmetries} in the two phases.

Liquid and crystal are just two examples. In fact, particles can organize in
many ways which lead to many different orders and many different types of
materials.  Landau symmetry-breaking theory\cite{L3726,GL5064,LanL58} provides
a general and a systematic understanding of these different orders.  It points
out that different orders really correspond to different symmetries in the
organizations of the constituent particles.  As a material changes from one
order to another order (\ie, as the material undergoes a phase transition),
what happens is that the symmetry of the organization of the particles changes.
Landau symmetry-breaking theory is a very successful theory. For a long time,
physicists believed that Landau symmetry-breaking theory describes all possible
orders in materials, and all possible (continuous) phase transitions.

\subsection{Quantum phases of matter}

Quantum phases of matter are phases of matter at zero temperature.  So quantum
phases correspond to the ground states of the quantum Hamiltonians that govern
the systems.  In this article, we will mainly discuss those quantum phases of
matter.  Crystal, conductor, insulator, superfluid, and magnets can exist at
zero temperature and are examples of quantum phases of matter.  

Again, physicists used to believe that Landau symmetry-breaking
theory also describes all possible quantum phases of matter, and all possible
(continuous) quantum phase transitions.  (Quantum phase transitions are zero
temperature  phase transitions.) For example, the superfluid is described by a
$U(1)$ symmetry breaking.  

\rem{It is interesting to compare a finite-temperature phase, \emph{liquid},
with a zero-temperature phase, \emph{superfluid}.  A liquid is described a
random \emph{probability} distributions of particles (such as atoms), while a
superfluid is described by a quantum wave function which is the
\emph{superposition} of a set of random particle configurations:
\begin{align}
 |\Phi_\text{superfluid}\>=
\sum_\text{random configurations}
\left  | 
\bmm \includegraphics[scale=0.25]{liquid} \emm 
\right  \>
\end{align}
The superposition of many different particle positions are called quantum
fluctuations in particle positions.

Since Landau symmetry-breaking theory suggests that all quantum phases are
described by symmetry breaking, thus we can use group theory to classify all
those symmetry breaking phases: All symmetry breaking quantum phases are
classified by a pair of mathematical objects $(G_H,G_\Phi)$, where $G_H$ is the
symmetry group of the Hamiltonian and $G_\Phi$ is the symmetry group of the
ground state.  For example,  a boson superfluid is labeled by $(U(1),\{1\})$,
where $U(1)$ is the symmetry group of the boson Hamiltonian which conserve the
boson number, and $\{1\}$ is the trivial group that contains only identity.  }

\section{Topological order}

\subsection{The discovery of topological order}

However, in late 1980s, it became clear that Landau symmetry-breaking theory
did not describe all possible phases.  In an attempt to explain high
temperature superconductivity, the chiral spin state was
introduced.\cite{KL8795,WWZ8913} At first, physicists still wanted to use
Landau symmetry-breaking theory to describe the chiral spin state. They
identified the chiral spin state as a state that breaks the time reversal and
parity symmetries, but not the spin rotation symmetry.\cite{WWZ8913}  This
should be the end of story according to Landau symmetry breaking description of
orders.  

But, it was quickly realized that there are many different chiral spin states
that have exactly the same symmetry.\cite{Wtop} So symmetry alone was not
enough to characterize and distinguish different chiral spin states.  This means
that the chiral spin states must contain a new kind of order that is beyond the
usual symmetry description.  The proposed new kind of order was named
``topological order''.\cite{Wrig} (The name ``topological order'' was motivated
by the low energy effective theory of the chiral spin states which is a
Chern-Simons theory\cite{WWZ8913} -- a topological quantum field theory
(TQFT)\cite{W8951}).  New quantum numbers (or new topological probes), such as
ground state degeneracy\cite{Wtop,WNtop}  and the non-Abelian geometric phase
of degenerate ground states\cite{Wrig,KW9327}, were introduced to
characterize/define the different topological orders in chiral spin states.  
%Recently, it was shown that topological orders can also be characterized by
%topological entropy.\cite{LW0605,KP0604}

\begin{figure}[tb]
\centerline{
\hfil
$\bmm
\includegraphics[height=1.7in]{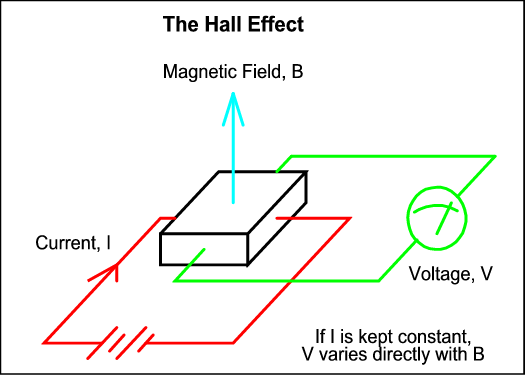}
\emm $
\hfil
$\bmm
\includegraphics[scale=0.5]{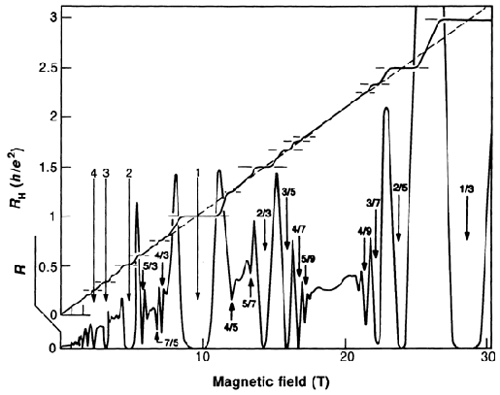}
\emm $
}
\caption{ 
2D electrons in strong magnetic field may form FQH states.  Each FQH state has
a quantized Hall coefficient $R_H$.
}
\label{Hall}
\end{figure}

But experiments soon indicated that chiral spin states do not describe
high-temperature superconductors, and the theory of topological order became a
theory with no experimental realization. However, the similarity\cite{KL8795}
between chiral spin states and fractional quantum Hall (FQH)
states\cite{TSG8259,L8395} allows one to use the theory of topological order to
describe different FQH states.

FQH states are gapped ground states of 2D electrons under strong magnetic
field.  FQH states have a property that a current density will induce an
electron field in the transverse direction: $E_y=R_H j_x$ (see Fig.
\ref{Hall}).  It is an amazing discovery that the Hall coefficient $R_H$ of a
FQH state is precisely quantized as a rational number $\frac{p}{q}$ if we
measure the Hall coefficient $R_H$ in unit of $\frac{h}{e^2}$: $
R_H=\frac{p}{q}\frac{h}{e^2} $ (see Fig. \ref{Hall}).\cite{TSG8259}  Different
quantized $R_H$ correspond to different FQH states. Just like the chiral spin
states, different FQH states all have the same symmetry and cannot be
distinguished by symmetry-breaking.  So there is no way to use different
symmetry breaking to describe different FQH states, and FQH states must contain
\emph{new} orders.  One finds that the new orders in quantum Hall states can
indeed be described by topological orders.\cite{WNtop} So the topological order
does have experimental realizations.

\rem{ We would like to point out that before the topological-order
understanding of FQH states, people have tried to use the notions of
off-diagonal long-range order and order parameter from Ginzburg-Landau theory
to describe FQH states.\cite{GM8752,R8986,ZHK8982,EI9137} Such an effort 
leads to a Ginzburg-Landau Chern-Simons effective theory for FQH
states.\cite{ZHK8982,EI9137} At same time, it was also realized that the  order
parameter in the Ginzburg-Landau Chern-Simons is not gauge invariant and is not
physical.  This is consistent with the topological-order understanding of FQH
states which suggests that FQH has no off-diagonal long-range order and cannot
be described by local order parameters.  So we can use effective theories
without order parameters to describe FQH states, and such  effective theories
are pure Chern-Simons effective
theories.\cite{WNtop,BW9045,FZ9117,FK9169,WZ9290,FS9333} The pure Chern-Simons
effective theories lead to a K-matrix classification\cite{WZ9290} of all
Abelian topologically ordered states (which include all Abelian FQH states).

FQH states were discovered in 1982\cite{TSG8259} before the introduction of the
concept of topological order.  But FQH states are not the first experimentally
discovered topologically ordered states.  The real-life superconductors, having
a $Z_2$ topological order,\cite{W9141,Wsrvb,HOS0497} were first experimentally
discovered topologically ordered states.\footnote{Note that real-life
superconductivity can be described by the Ginzburg-Landau theory with a
\emph{dynamical} $U(1)$ gauge field, which become a $Z_2$ gauge theory at low energies; and  a
$Z_2$ gauge theory is an effective theory of $Z_2$ topological order. 
%The prediction of the vortex state in superconductors was one of the main
%successes of Ginzburg-Landau theory with dynamical U(1) gauge field. The
%vortex in the gauged Ginzburg-Landau theory is nothing but the Z2 flux line in
%the Z2 gauge theory. 
In many textbook, superconductivity is described by the Ginzburg-Landau theory
without the dynamical $U(1)$ gauge field, which fails to describe the real-life
superconductors with dynamical electromagnetic interaction.
Such a textbook superconductivity is described by a $U(1)$ symmetry breaking.} 
(Ironically, the
Ginzburg-Landau symmetry breaking theory was developed to describe
superconductors, despite the real-life superconductors are not symmetry
breaking states, but topologically ordered states.) }

\subsection{Intuitive pictures of topological order}

Topological order is a very new concept that describes quantum entanglements in
many-body systems.  Such a concept is very remote from our daily experiences
and it is hard to have an intuition about it.  So before we define topological
order in general terms (which can be abstract), let us first introduce and
explain the concept through some intuitive pictures.

\begin{figure}[tb]
\hfil
\includegraphics[height=1.1in]{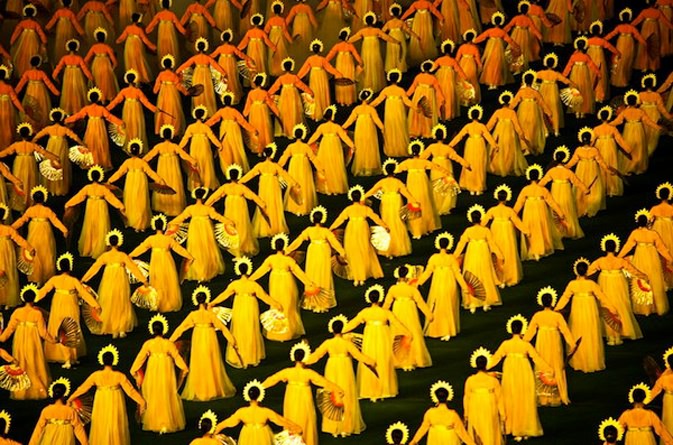}
\hfil
\includegraphics[height=1.1in]{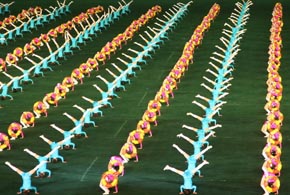}\\
\hfil
Ferromagnet 
\hfil
~~~~~ Anti-ferromagnet\\[2mm]
\hfil
\includegraphics[height=1.1in]{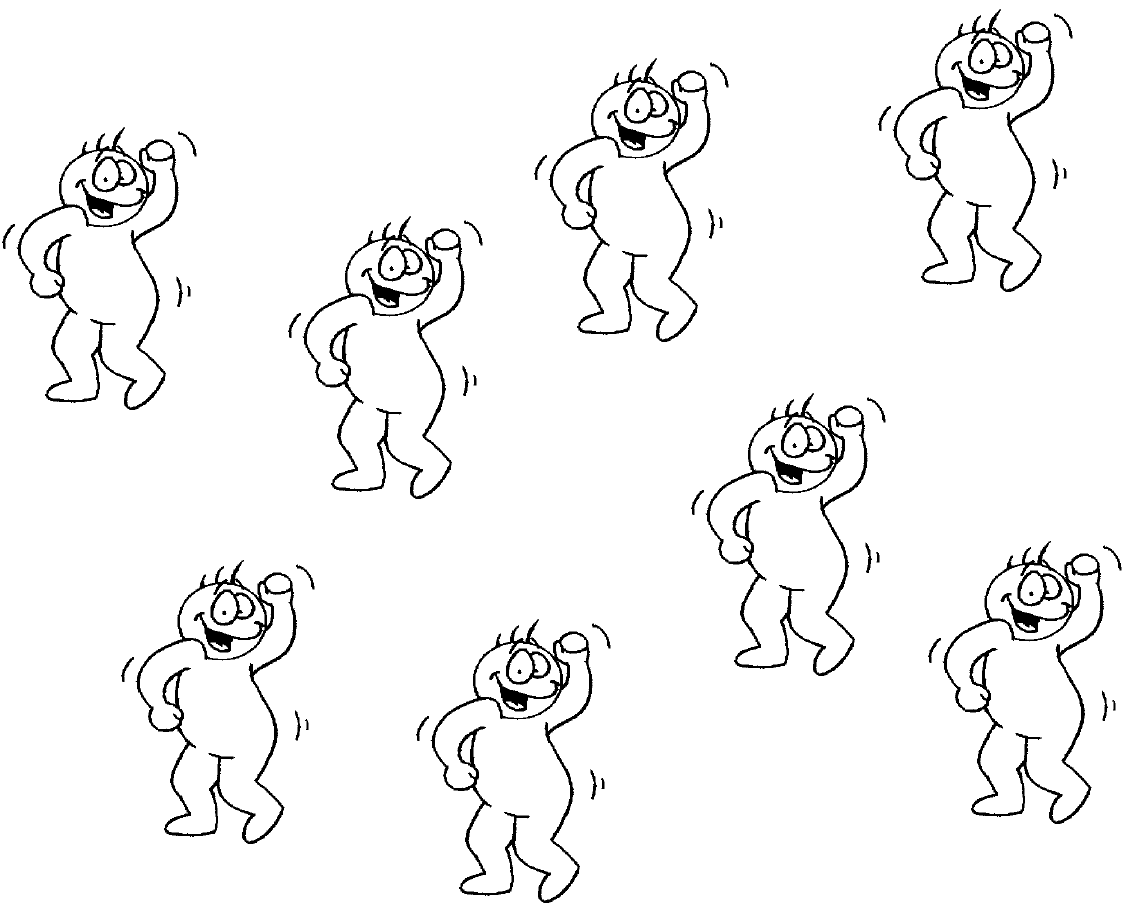}
\hfil
\hfil
\includegraphics[height=1.1in]{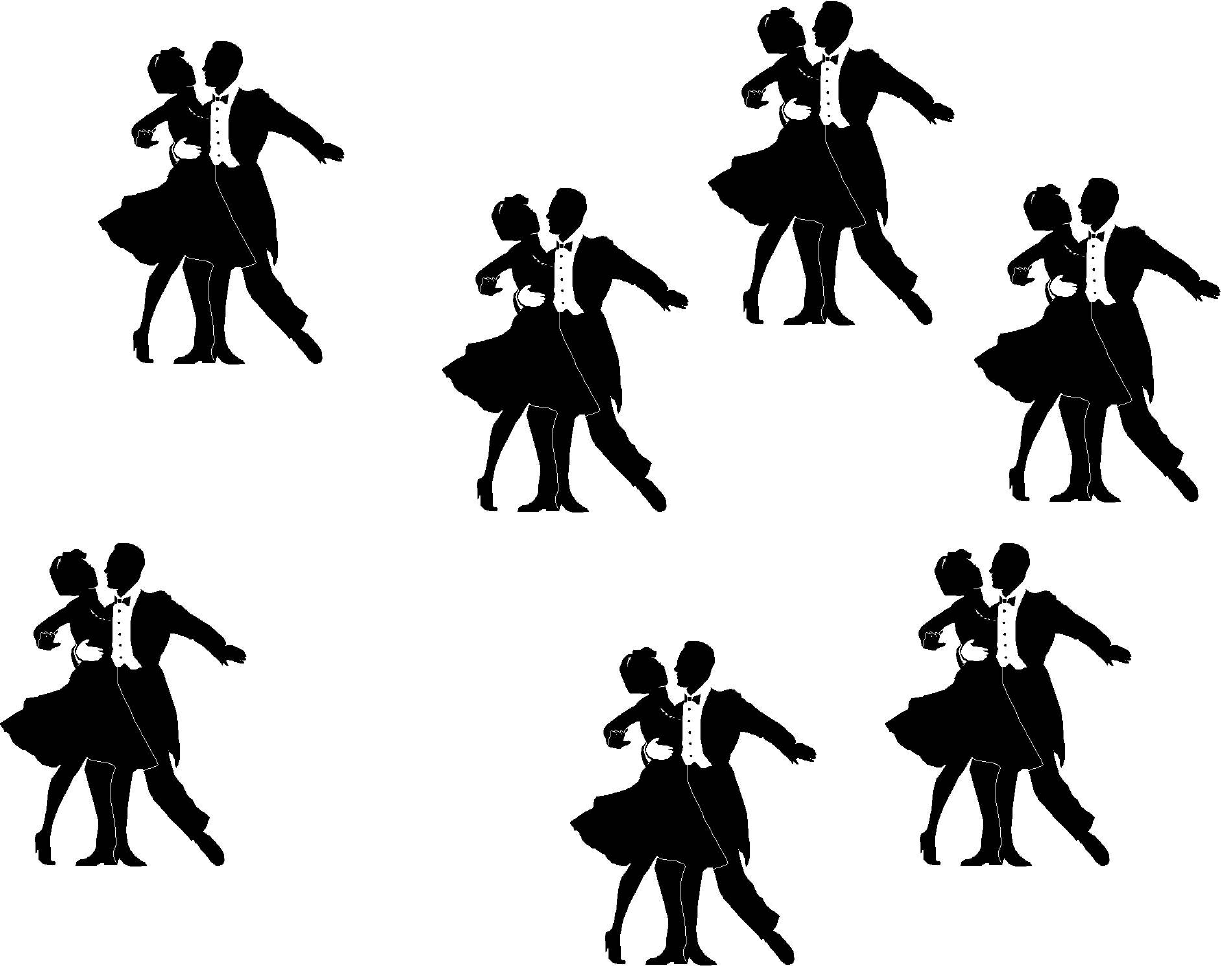}\\
\hfil
Superfluid of bosons
\hfil
~~~~ Superfluid of fermions
\caption{
The dancing patterns for
the symmetry breaking orders.
}
\label{symmpic}
\end{figure}

We can use dancing to gain an intuitive picture of topological order. But
before we do that, let us use dancing picture to describe the old symmetry
breaking orders (see Fig.  \ref{symmpic}).  In the symmetry breaking orders,
every particle/spin (or every pair of particles/spins) dance by itself, and
they all dance in the same way.  (The ``same way'' of dancing represents a
long-range order.)  For example, in a ferromagnet, every electron has a fixed
position and the same spin direction.  We can describe an anti-ferromagnet by
saying every pair of electrons has a fixed position and the two electrons in a
pair have opposite spin directions.  In a boson superfluid, each boson is
moving around by itself and doing the same dance, while in a fermion
superfluid, fermions dance around in pairs and each pair is doing the same
dance.

\begin{figure}[tb]
\hfil
\includegraphics[height=1.2in]{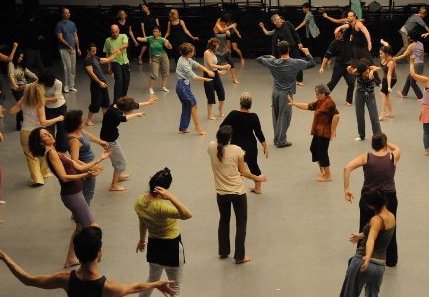}
\hfil
\hfil
\includegraphics[height=1.2in]{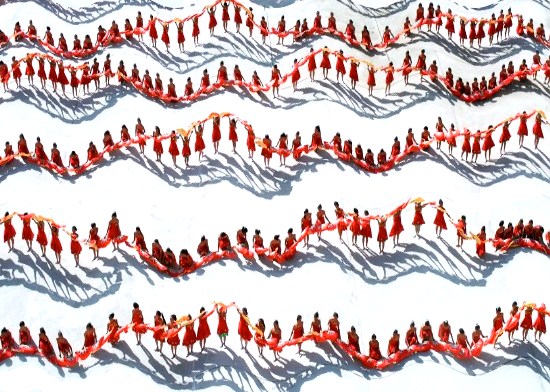}\\
\hfil
~~~~~ FQH state 
\hfil
\hfil
~~~~~ 
~~~~~ 
String liquid (spin liquid)
\caption{
The dancing patterns for the topological orders.
}
\label{toppic}
\end{figure}

We can also understand topological orders through such dancing pictures.
Unlike fermion superfluid where fermion dance in pairs, a topological order is
described by a global dance, where every particle is dancing with every other
particles in a very organized way:  (a) all spins/particles dance following a
set of \emph{local} dancing ``rules'' trying to lower the energy of a
\emph{local} Hamiltonian.  (b) If all the spins/particles follow the local
dancing ``rules'', then they will form a global dancing pattern, which
correspond to the topological order.  
%We see that topological order is a pattern of collective dancing of all
%particles/spins (in contrast, fermions dance in pairs in a fermion
%superfluid).
(c) Such a global pattern of collective dancing is a pattern of quantum
fluctuation which corresponds to a pattern of \emph{long range entanglements}.

For example in FQH liquid, 
the electrons dance following the following local
dancing rules:\\ 
(a) electron always dances anti-clockwise which implies that
the electron wave function only depend on the electron coordinates $(x,y)$
via $z=x+\imth y$.\\
(b) each electron always takes exact three steps to dance around any
other electron,
which implies that the phase of the
wave function changes by $6\pi$ as we move an electron around
any other electron.\\
The above two local dancing rules fix a global dance pattern which correspond
to the Laughlin wave function\cite{L8395} $\Phi_\text{FQH} = \prod
(z_i-z_j)^3$.  Such an collective dancing gives rise to the
topological order (or long range entanglements) in the FQH state.

\begin{figure}[tb]
\centerline{
\includegraphics[height=1.2in]{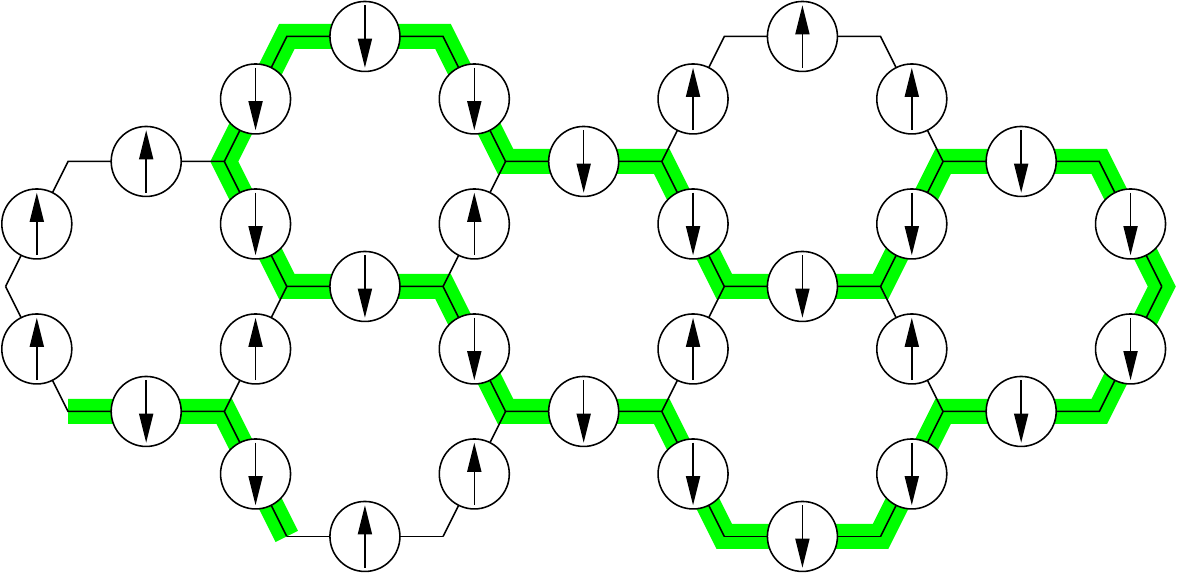}
}
\caption{
The strings in a spin-1/2 model.
In the background of up-spins, the down-spins form closed strings.
}
\label{strspin}
\end{figure}

\begin{figure}[tb]
\centerline{
\includegraphics[height=1.in]{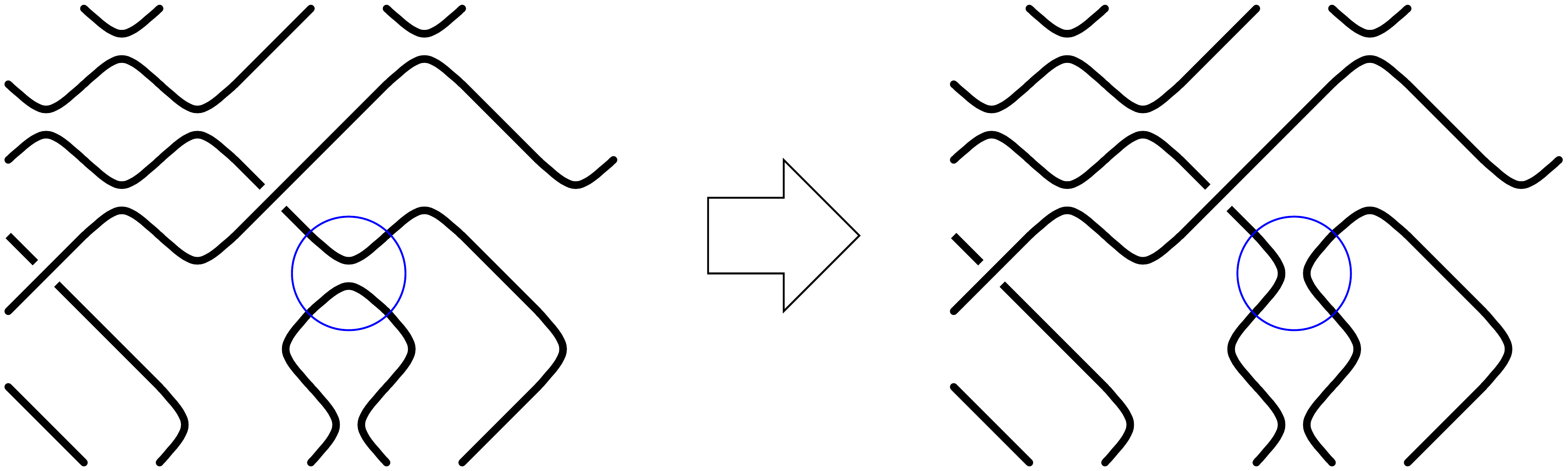}
}
\caption{
In string liquid, strings can move freely,
including reconnecting the strings.
}
\label{strnetSa}
\end{figure}

In additional to FQH states, some spin liquids also contain topological
orders.\cite{WWZ8913,RS9173,Wsrvb,MLB9964,MS0181} 
(Spin liquids refer to ground states of quantum spin systems that do not
break any symmetry of the spin Hamiltonian.)
In those spin liquids, the
spins ``dance'' following the follow local dancing rules:\\
(a) Down spins form
closed strings with no ends in the background of up-spins (see Fig.  \ref{strspin}).\\
(b) Strings can otherwise move freely, including reconnecting freely (see Fig.
\ref{strnetSa}). \\ 
The global dance formed by the spins following the above dancing rules gives us
a quantum spin liquid which is a superposition
of all closed-string configurations:\cite{K032}
$|\Phi_\text{string}\>=\sum_\text{all string pattern} \left |\bmm
\includegraphics[height=0.3in]{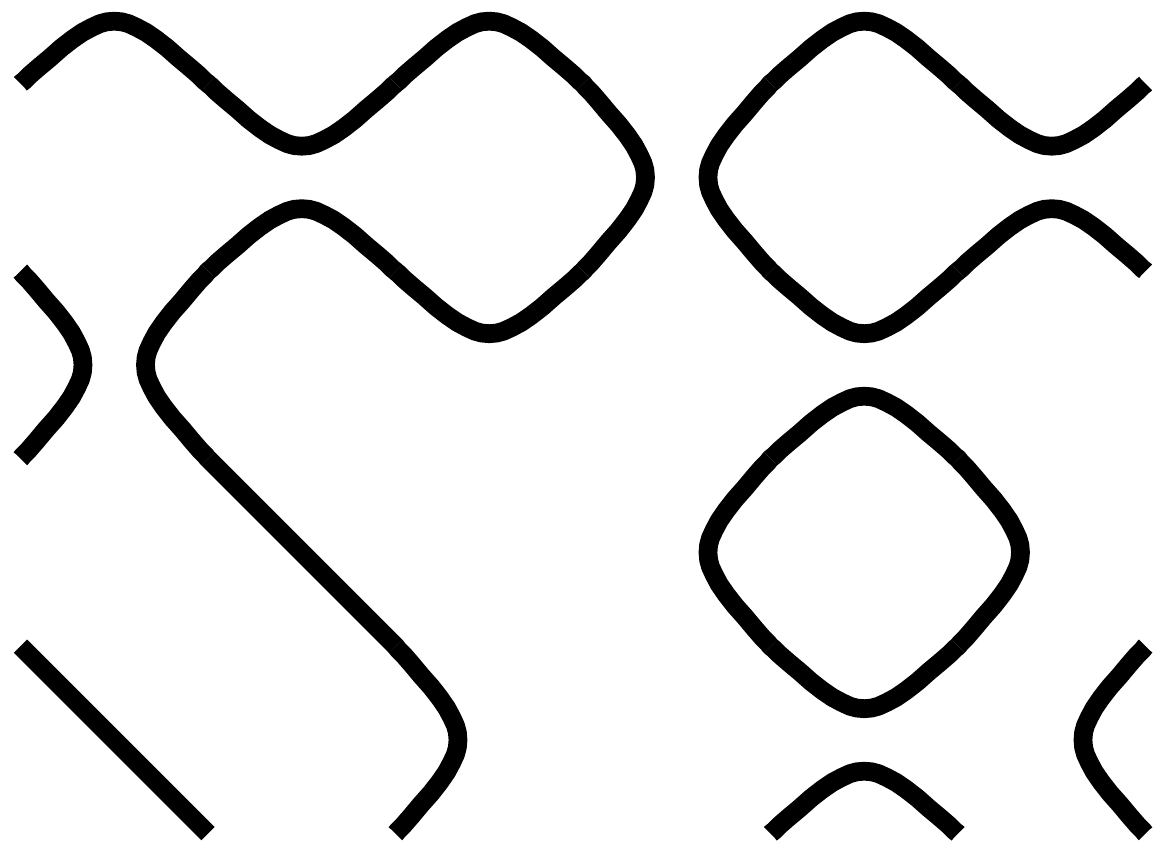}\emm\right \> $.  Such a state is called
a string or string-net condensed state.\cite{LWstrnet} The collective dancing
gives rise to a non-trivial topological order and a pattern of long range
entanglements in the spin liquid state.

\section{What is the significance of topological order?}

%Although the topological degeneracy and the non-Abelian geometric phases of
%the degenerate ground states (see Fig. \ref{g0g1g2} and \ref{modtrns}) are
%experimental probes that can systematically probe topological order (or
%long-range entanglements).  However, these knids of probes are very abstract
%and can hardly be realized in experiments.  

The above descriptions of topological order is intuitive and not concrete. It
is not clear if the topological order (the global dancing pattern or the
long-range entanglement) has any experimental significance.  In order for the
topological order to be a useful concept, it must have new experimental
properties that are different from any symmetry breaking states. Those new
experimental properties should indicate the non-trivialness of the topological
order.  In fact, the concept of topological order should be defined by the
collection of those new experimental properties.  

Indeed, topological order does have new  characteristic properties.  Those
properties of topological orders reflect the significance of topological order:

%So one may ask: Are there easier and more practical ways to measure/study
%topological order (or long-range entanglements)?  What is the experimental
%significance of topological order?  

%In fact, there are indeed new characteristic properties of topological orders
%(or long-range entanglements) that can be measured in practical experiments.

\noindent (1) Topological orders produce new kind of waves (\ie the collective
excitations above the topologically ordered ground
states).\cite{Wlight,SM0204,Walight,Wqoem,MS0312,HFB0404,LWuni,LWqed,CMS1235}  The new
kind of waves can be probed/studied in practical experiments, such as neutron
scattering experiments.\cite{MS0312}

\noindent (2) The finite-energy defects of topological order (\ie the
quasiparticles) can carry fractional statistics\cite{H8483,ASW8422} (including
non-Abelian statistics\cite{Wnab,MR9162}) and fractional
charges\cite{JR7698,L8395} (if there is a symmetry).  Such a property allows us
to use topologically ordered states as a medium for topological quantum
memory\cite{DKL0252} and topological quantum computations.\cite{K032}

\noindent (3) Some  topological orders have topologically protected gapless
boundary excitations.\cite{H8285,Wedge,M9020} Such gapless boundary excitations
are topologically protected, which lead to perfect conducting boundary channels
even with magnetic impurities.\cite{KDP8094}  This property may lead to device
applications.

In the following, we will study some examples of topological orders and reveal
their amazing topological properties.

\section{Examples of topological order: quantum liquid of oriented strings and
a unification of gauge interactions and Fermi statistics}

Our first example is a  quantum liquid of oriented strings.  We will discuss
its new topological properties (1) and (2).  We find that the new kind of waves
and the emergent statistics are so profound, that they may change our view of
universe. Let us start by explaining a basic notion -- ``principle of
emergence''.

\subsection{Principle of emergence} \label{pem}

Typically, one thinks the properties of a material should be determined by the
components that form the material.  However, this simple intuition is
incorrect, since all the materials are made of same components: electrons,
protons and neutrons.  So we cannot use the richness of the components to
understand the richness of the materials. In fact, the various properties of
different materials originate from various ways in which the particles are
organized.  The organizations of the particles are called orders.  Different
orders (the organizations of particles) give rise to different physical
properties of a material. It is richness of the orders that gives rise to the
richness of material world.

%From very early on, people realized that two kinds of matter, solids and
%liquids, have very different mechanical properties.  Basically, a liquid has
%no shape and can flow freely.  In contrast, a solid has a fixed shape and can
%retain its shape.  Using a more technical term, we say that a solid can resist
%shear deformations while a liquid cannot.

\begin{figure}[t]
\centerline{
\includegraphics[scale=0.8]{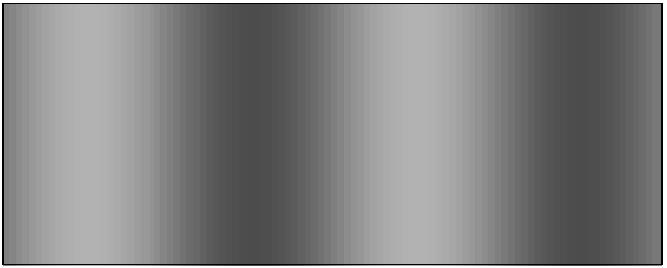}
}
\caption{
Liquids only have a compression wave -- a wave of density fluctuations.
}
\label{liquidC}
\end{figure}

Let us use the origin of mechanical properties and the origin of waves to
explain, in a more concrete way, how orders determine the physics properties of
a material.  We know that a deformation in a material can propagate just like
the ripple on the surface of water.  The propagating deformation corresponds to
a wave traveling through the material.  Since liquids can resist only
compression deformation, so liquids can only support a single kind of wave --
compression wave (see Fig.  \ref{liquidC}). (Compression wave is also
called longitudinal wave.) Mathematically the motion of the compression wave is
governed by the Euler equation
%\footnote{To be precise, the liquids discussed in this article are liquids at
%zero temperature. The zero-temperature liquids are always superfluids.}
\begin{equation}
\label{EulEq}
 \frac{\prt^2 \rho}{\prt t^2}-v^2 \frac{\prt^2 \rho}{\prt x^2}=0,
\end{equation}
where $\rho$ is the density of the liquid.  

Solid can resist both compression
and shear deformations.  As a result, solids can support both compression wave
and transverse wave. The transverse wave correspond to the propagation of shear
deformations. In fact there are two transverse waves corresponding to two
directions of shear deformations.  The propagation of the compression wave and
the two transverse waves in solids are described by the elasticity equation
\begin{equation}
\label{NavEq}
 \frac{\prt^2 u^i}{\prt t^2}- T^{ikl}_j \frac{\prt^2 u^j}{\prt x^k\prt x^l}
=0
\end{equation}
where the vector field $u^i(\v x, t)$  describes the local displacement of the
solid. 

\begin{figure}[t]
\centerline{
\includegraphics[scale=0.42]{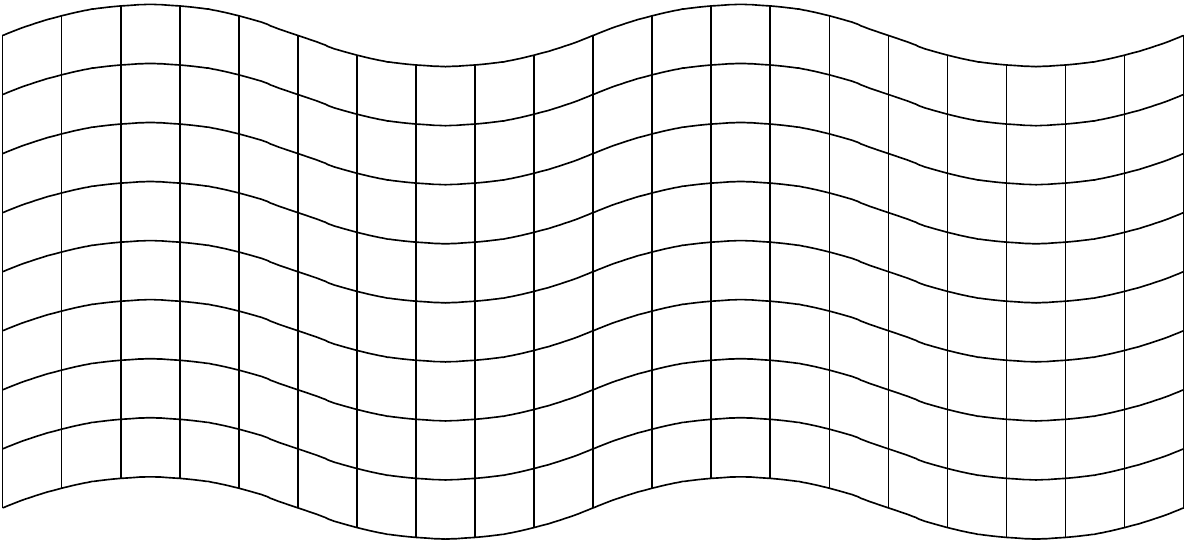}
}
\caption{
Drawing a grid on a sold helps us to see the deformation of the solid.  The
vector $u^i$ in \eqn{NavEq} is the displacement of a vertex in the grid.  In
addition to the compression wave (\ie the density wave), a solid also supports
transverse wave (wave of shear deformation) as shown in the above figure.
}
\label{crystalC}
\end{figure}

We would like to point out that the elasticity equation and the Euler equations
not only describe the propagation of waves, they actually describe all small
deformations in solids and liquids.  Thus, the two equations represent a
complete mathematical description of the mechanical properties of solids and
liquids.  

But why do solids and liquids behave so differently? What makes a solid to have
a shape and a liquid to have no shape?  What are the origins of
elasticity  equation and Euler equations? The answer to those questions has to
wait until the discovery of atoms in 19th century.  Since then, we realized
that both solids and liquids are formed by collections of atoms. The main
difference between the solids and liquids is that the atoms are organized very
differently.  In liquids, the positions of atoms fluctuate randomly (see Fig.
\ref{cryliq1}a),  while in solids, atoms organize into a regular fixed array
(see Fig.  \ref{cryliq1}b).\footnote{The solids here should be more accurately
referred as crystals.}  It is the different organizations of atoms that lead to
the different mechanical properties of liquids and solids.  In other words, it
is the different organizations of atoms that make liquids to be able to flow
freely and solids to be able to retain its shape.

\begin{figure}[t]
\centerline{
\includegraphics[scale=0.5]{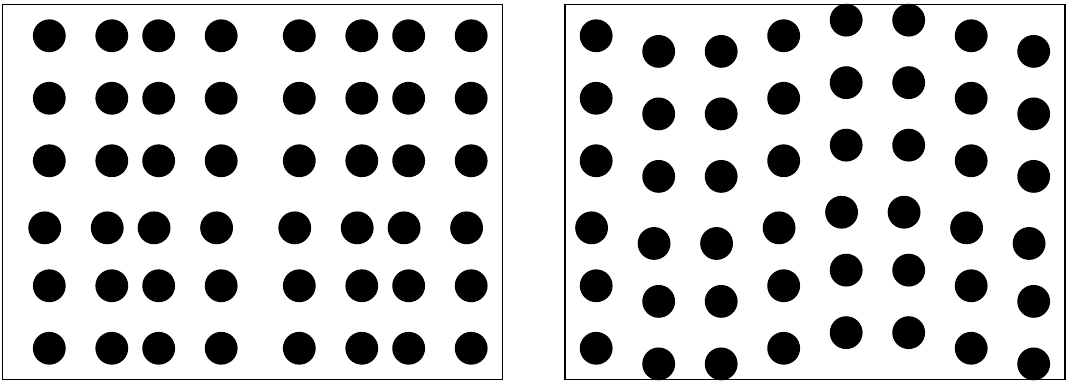}
}
\centerline{
\hfil
(a)
\hfil
(b)
\hfil
}
\caption{
The atomic picture of 
(a) the compression wave and
(b) the transverse wave in a crystal.
}
\label{sndwavs1}
\end{figure}

\begin{figure}[t]
\centerline{
\includegraphics[scale=0.5]{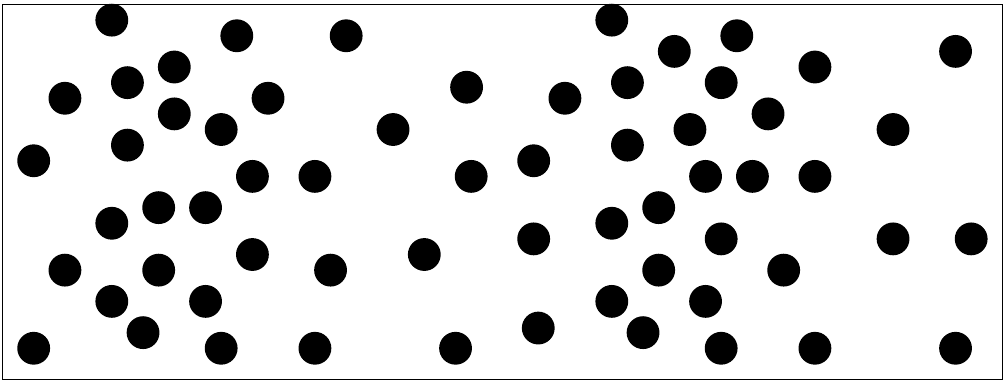}
}
\caption{
The atomic picture of the compression wave in liquids. 
}
\label{liquidwav1}
\end{figure}

How can different organizations of atoms affect mechanical properties of
materials?  In solids, both the compression deformation (see Fig.
\ref{sndwavs1}a) and the shear deformation (see Fig. \ref{sndwavs1}b) lead to
real physical changes of the atomic configurations. Such changes cost energies.
As a result, solids can resist both kinds of deformations and can retain their
shapes.  This is why we have both the compression wave and the transverse wave
in solids.

In contrast, a shear deformation of atoms in liquids does not result in a new
configuration since the atoms still have uniformly random positions. So the
shear deformation is a do-nothing operation for liquids.  Only the compression
deformation which changes the density of the atoms results in a new atomic
configuration and costs energies. As a result, liquids can only resist
compression and have only compression wave.  Since shear deformations do not
cost any energy for liquids, liquids can flow freely.

We see that the properties of the propagating wave are entirely determined by
how the atoms are organized in the materials.  Different organizations lead to
different kinds of waves and different kinds of mechanical laws.  Such a point
of view of different kinds of waves/laws originated from different
organizations of 
%possibly the same kind of 
particles is a central theme in condensed matter physics.  This point of view
is called the principle of emergence.  

\subsection{String-net liquid unifies light and electrons}

The elasticity equation and the Euler equation are two very important
equations.  They lay the foundation of many branches of science such as
mechanical engineering, aerodynamic engineering, \etc.  But, we have a more
important equation, Maxwell equation, that describes light waves in vacuum.
When Maxwell equation was first introduced, people firmly believed that any
wave must corresponds to motion of something.  So people want to find out
what is the origin of the Maxwell equation?  The motion of what gives rise
electromagnetic wave?

First, one may wonder: can Maxwell equation comes from a certain symmetry
breaking order?  Based on Landau symmetry-breaking theory, the different
symmetry breaking orders can indeed lead to different waves satisfying
different wave equations.
% that correspond to
%the fluctuations of order parameter in the symmetry breaking state.  
So maybe a certain symmetry breaking order can give rise to a wave that satisfy
Maxwell equation.  But people have been searching for ether -- a medium that
supports light wave -- for over 100 years, and could not find any symmetry
breaking states that can give rise to waves satisfying the Maxwell equation.
This is one of the reasons why people give up the idea of ether as the origin
of light and Maxwell equation.

However, the discovery of topological order\cite{Wtop,Wrig} suggests that
Landau symmetry-breaking theory does not describe all possible organizations of
bosons/spins.  This gives us a new hope: Maxwell equation may arise from a new
kind of organizations of bosons/spins that have non-trivial topological orders.

In addition to the Maxwell equation, there is an even stranger equation, Dirac
equation, that describes wave of electrons (and other fermions).  Electrons
have Fermi statistics. They are fundamentally different from the quanta of
other familiar waves, such as photons and phonons, since those quanta all have
Bose statistics.  To describe the electron wave, the amplitude of the wave must
be anti-commuting Grassmann numbers, so that the wave quanta will have Fermi
statistics. Since electrons are so strange, few people regard electrons and the
electron waves as collective motions of something. People accept without
questioning that electrons are fundamental particles, one of the building
blocks of all that exist.  

However, from a condensed matter physics point of
view, all low energy excitations are collective motion of something.  If we try
to regard photons as collective modes, why cann't we regard electrons as
collective modes as well?  So maybe, Dirac equation and the associated fermions
can also arise from a new kind of organizations of bosons/spins that have
non-trivial topological orders.

\begin{figure}[tb]
\centerline{
\includegraphics[scale=0.15]{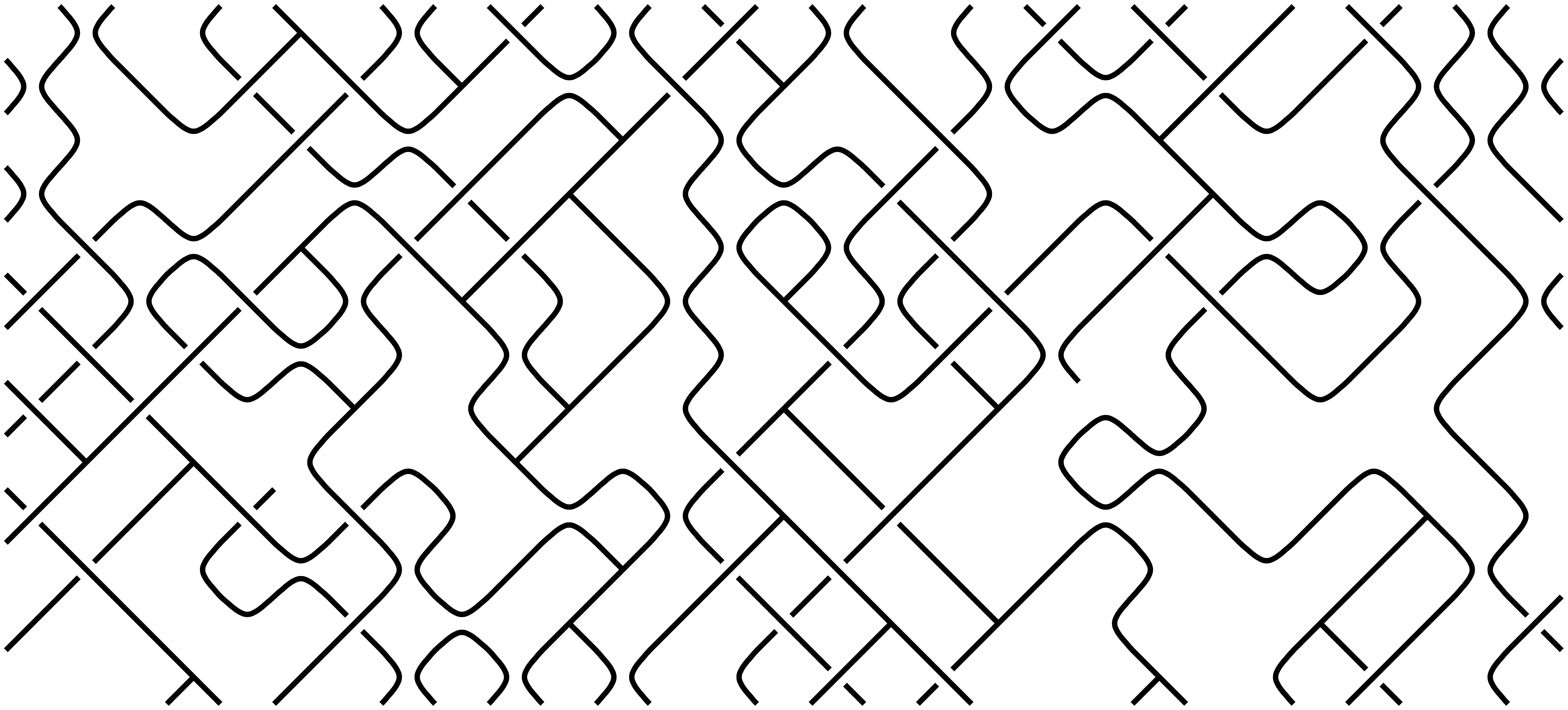}
}
\caption{
A quantum ether: The fluctuation of oriented strings give rise to
electromagnetic waves (or light). The ends of strings give rise to electrons.
Note that oriented strings have directions which should be described by curves
with arrow. For ease of drawing, the arrows on the curves are omitted in the
above plot.
}
\label{stringnetS}
\end{figure}

A recent study provides an positive answer to the above
questions.\cite{LWstrnet,LWuni,LWqed} We find that if bosons/spins form large
oriented strings and if those strings form a quantum liquid state, then the
collective motion of the such organized bosons/spins will correspond to waves
described by Maxwell equation and Dirac equation.  The strings in the string
liquid are free to join and cross each other. As a result, the strings look
more like a network (see Fig.  \ref{stringnetS}).  For this reason, the string
liquid is actually a liquid of string-nets, which is called string-net
condensed state.

\begin{figure}[tb]
\centerline{
\includegraphics[width=2.5in]{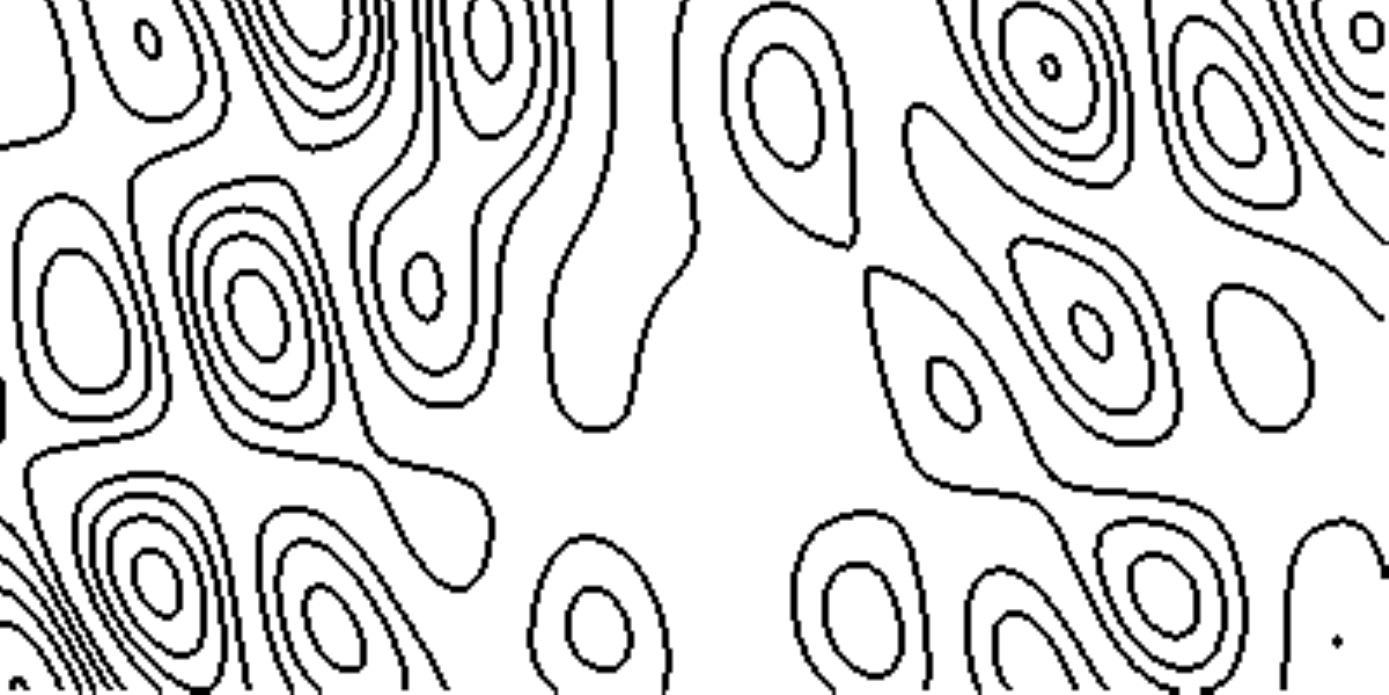}
}
\caption{
The fluctuating strings in a string liquid.
%A quantum ether: The fluctuation of oriented strings give rise to
%electromagnetic waves (or light). The ends of strings give rise to electrons.
%Note that oriented strings have directions which should be described by curves
%with arrow. For ease of drawing, the arrows on the curves are omitted in the
%above plot.
}
\label{vacBW}
\end{figure}

\begin{figure}[tb]
\centerline{
\includegraphics[width=2.5in]{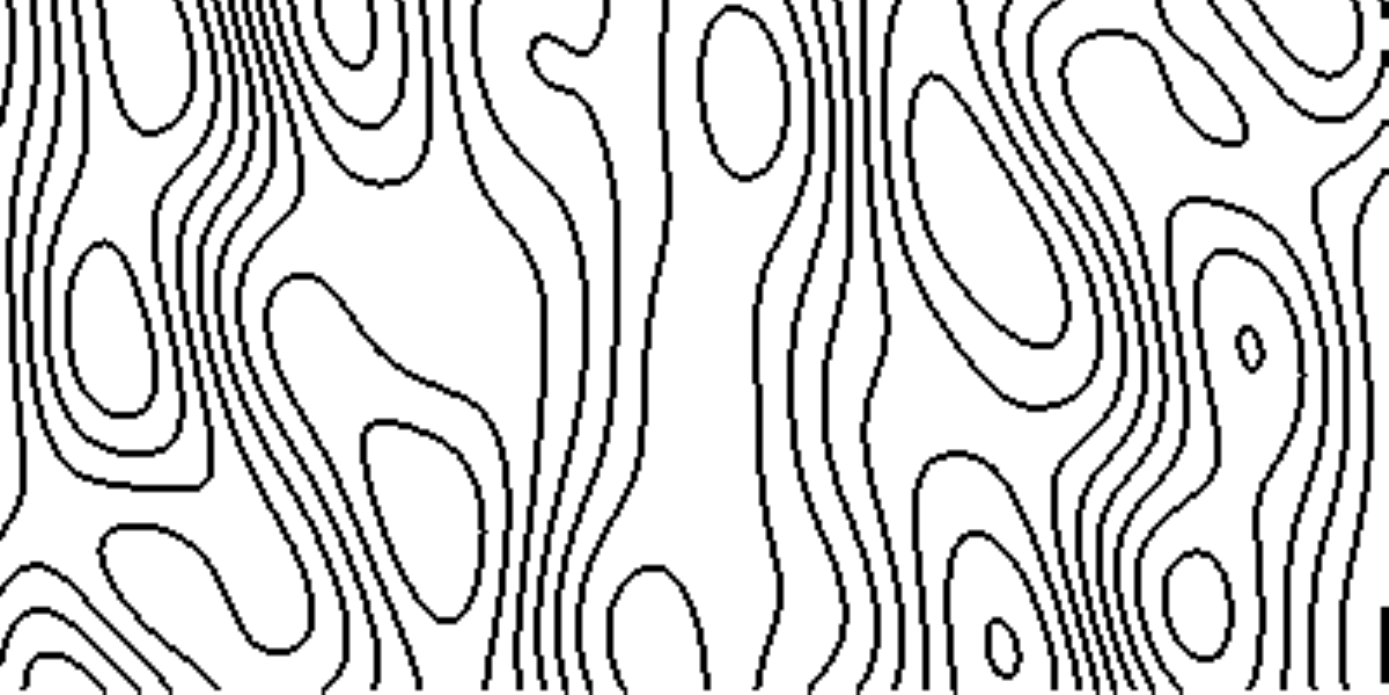}
}
\caption{
A ``density'' wave of oriented strings in a string liquid.  The wave propagates
in $\v x$-direction.  The ``density'' vector $\v E$ points in $\v y$-direction.
For ease of drawing, the arrows on the oriented strings are omitted in the
above plot.
}
\label{vacwvBW}
\end{figure}

But why the waving of strings produces waves described by the Maxwell equation?
We know that the particles in a liquid have a random but uniform distribution.
A deformation of such a distribution corresponds a density fluctuation, which
can be described by a scaler field $\rho(\v x,t)$.  Thus the waves in a liquid
is described by the scaler field $\rho(\v x,t)$ which satisfy the Euler
equation \eq{EulEq}.  Similarly, the strings in a string-net liquid also have
a random but uniform distribution (see Fig.  \ref{vacBW}). A deformation of
string-net liquid corresponds to a change of the density of the strings (see
Fig. \ref{vacwvBW}).  However, since strings have an orientation, the
``density'' fluctuations are described by a vector field $\v E(\v x, t)$, which
indicates there are more strings in the $\v E$ direction on average.  The
oriented strings can be regarded as flux lines. The vector field $\v E(\v x,
t)$ describes the smeared average flux.  
%In fact $\v E(\v x, t)$ correspond to the electric field in the
%electromagnetic wave.} The field $\v E(\v x, t)$ can only describe transverse
%waves due to the flux conservation $\v\prt\cdot\v E(\v x, t)=0$.}  
Since strings are continuous (\ie they cannot end), the
flux is conserved: $\v\prt\cdot\v E(\v x, t)=0$.  The vector density $\v E(\v
x, t)$ of strings cannot change in the direction along the strings (\ie along
the $\v E(\v x, t)$ direction). $\v E(\v x, t)$ can change only in the
direction perpendicular to $\v E(\v x, t)$.  Since the direction of the
propagation is the same as the direction in which $\v E(\v x, t)$ varies, thus
the waves described by $\v E(\v x, t)$ must be transverse waves: $\v E(\v x,
t)$ is always perpendicular to the direction of the propagation.  Therefore,
the waves in the string liquid have a very special property: the waves have
only transverse modes and no longitudinal mode.  This is exactly the property
of the light waves described by the Maxwell equation.  We see that ``density''
fluctuations of strings (which are described be a transverse vector field)
naturally give rise to the light (or electromagnetic) waves and the Maxwell
equation.\cite{Walight,Wqoem,MS0312,HFB0404,LWuni,LWqed}

It is interesting to compare solid, liquid, and string-net liquid.
We know that the particles in a solid organized into a regular
lattice pattern.  The waving of such organized particles produces a compression
wave and two transverse waves.  The particles in a liquid have a more random
organization.  As a result, the waves in liquids lost two transverse modes and
contain only a single compression mode.  The particles in a string-net liquid
also have a random organization, but in a different way.  The particles first
form string-nets and string-nets then form a random liquid state. Due to this
different kind of randomness, the waves in string-net condensed state lost the
compression mode and contain two transverse modes.  Such a wave (having only
two transverse modes) is exactly the electromagnetic wave.
% described by the Maxwell equation.

To understand how electrons appear from string-nets, we would like to point out
that if we only want photons and no other particles, the strings must be closed
strings with no ends.  The fluctuations of closed strings produce only photons.
If strings have open ends, those open ends can move around and just behave like
independent particles.  Those particles are not photons. In fact, the ends of
strings are nothing but electrons.

How do we know that ends of strings behave like electrons?  First, since the
waving of string-nets is an electromagnetic wave, a deformation of string-nets
correspond to an electromagnetic field.  So we can study how an end of a string
interacts with a deformation of string-nets.  We find that such an interaction
is just like the interaction between a charged electron and an electromagnetic
field. Also electrons have a subtle but very important property -- Fermi
statistics, which is a property that exists only quantum theory.  
%Without their Fermi statistics, the electrons in atoms would have a very
%different organization. All atoms would have very similar chemical properties
%and behave like a noble gas.  
Amazingly, the ends of strings can reproduce this subtle quantum property of
Fermi statistics.\cite{LWsta,LWstrnet}  Actually, string-net liquids explain
why Fermi statistics should exist.

We see that string-nets naturally explain both light and electrons (gauge
interactions and Fermi statistics).  In other words, string-net theory provides
a way to unify light and electrons.\cite{LWuni,LWqed} So, the fact that our
vacuum contains both light and electrons may not be a mere accident. It may
actually suggest that the vacuum is indeed a string-net liquid.  

\subsection{More general string-net liquid and emergence of non-Abelian gauge
theory}

Here, we would like to point out that there are many different kinds of
string-net liquids.  
%Strings in a string liquid can have several types.  
The strings in different liquids may have different numbers of types.  The
strings may also join in different ways.  For a general string-net liquid, the
waving of the strings may not correspond to light and the ends of strings may
not be electrons.  Only one kind of string-net liquids give rise to light and
electrons.  On the other hand, the fact that there are many different kinds of
string-net liquids allows us to explain more than just light and electrons.  We
can design a particular type of string-net liquids which not only gives rise to
electrons and photons, but also gives rise to quarks and
gluons.\cite{Wqoem,LWstrnet} The waving of such type of string-nets corresponds
to photons (light) and gluons. The ends of different types of strings
correspond to electrons and quarks. It would be interesting to see if it is
possible to design a string-net liquid that produces all elementary particles!
If this is possible, the ether formed by such string-nets can provide an origin
of all elementary particles.\footnote{So far we can use string-net to produce
almost all elementary particles, expect for the graviton that is responsible
for the gravity.  Also, we are unable to produce the chiral coupling between
the $SU(2)$ gauge boson and the fermions within the string-net picture.}

\rem{We like to stress that the particles that form the string-nets are bosons.
So in the string-net picture, both the Maxwell equation and Dirac equation,
emerge from \emph{local} bosonic models (or in other words, from local qubit
model).  

The electric field and the magnetic field in the Maxwell equation are called
gauge fields.  The field in the Dirac equation are Grassmann-number valued
field.\footnote{Grassmann numbers are anti-commuting numbers.} For a long time,
we though that we have to use gauge fields to describe light waves that have
only two transverse modes, and we though that we have to use Grassmann-number
valued fields to describe electrons and quarks that have Fermi statistics. So
gauge fields and Grassmann-number valued fields become the fundamental build
blocks of quantum field theory that describe our world.  The string-net liquids
demonstrate we do not have to introduce  gauge fields and  Grassmann-number
valued fields to describe photons, gluons, electrons, and quarks. It
demonstrates how gauge fields and Grassmann fields emerge from local bosonic
models (or from local qubit models) that contain only complex scaler fields at
the cut-off scale.  

Our attempt to understand light has long and evolving history.  We first
thought light to be a beam of particles.  After Maxwell, we understand light as
electromagnetic waves.  After Einstein's theory of general relativity, where
gravity is viewed as curvature in space-time, Weyl and others try to view
electromagnetic field as curvatures in the ``unit system'' that we used to
measure complex phases.  It leads to the notion of gauge theory.  The general
relativity and the gauge theory are two corner stones of modern physics. They
provide a unified understanding of all four interactions in terms of a
beautiful mathematical frame work: all interactions can be understood
geometrically as curvatures in space-time and in ``unit systems'' (or more
precisely, as curvatures in the tangent bundle and other vector bundles in
space-time).

Later, people in high-energy physics and in condensed matter physics have found
another way in which gauge field can emerge:\cite{DDL7863,W7985,BA8880,AM8874}
one first cut a particle (such as an electron) into two partons by writing the
field of the particle as the product of the two fields of the two partons.
Then one introduces a gauge field to glue the  two partons back to the original
particle.  Such a ``glue-picture'' of gauge fields (instead of the fiber
bundle picture  of gauge fields) allow us to understand the emergence of gauge
fields in models that originally contain no gauge field at the cut-off scale.

A string picture represent the third way to understand gauge theory.  String
operators appear in the Wilson-loop characterization\cite{W7445} of gauge
theory. The Hamiltonian and the duality description of lattice gauge theory
also reveal string structures.\cite{KS7595,BMK7793,K7959,S8053}.  Lattice gauge
theories are not local bosonic models and the strings are unbreakable in
lattice gauge theories.  String-net theory points out that even breakable
strings can give rise to gauge fields.\cite{HWcnt} So we do not really need
strings. Bosonic particles themselves are capable of generating gauge fields
and the associated Maxwell equation.  This phenomenon was discovered in several
bosonic models\cite{FNN8035,BA8880,Wlight,MS0204,HFB0404} before realizing
their connection to the string-net liquids.\cite{Walight}  Since gauge field
can emerge from local bosonic models (such as qubit models), the string picture
evolves into the entanglement picture -- the fourth way to understand gauge
field: gauge fields are fluctuations of long-range entanglements.  I feel that
the  entanglement picture capture the essence of gauge theory.  Despite the
beauty of the geometric picture, the essence of gauge theory is not the curved
fiber bundles.  In fact, we can view gauge theory as a theory for long-range
entanglements, although the gauge theory is discovered long before the notion
of long-range entanglements.

Viewing gauge field (and the associated gauge bosons) as fluctuations of
long-range entanglements has an added bonus: we can understand the origin of
Fermi statistics in the same way: fermions emerge are defects of long-range
entanglements, even though the original model is purely bosonic.  Previously,
there are two ways to obtain emergent fermions from purely bosonic model: by
binding gauge charge and gauge flux in (2+1)D,\cite{LM7701,W8257} and by
binding the charge and the monopole in a $U(1)$ gauge theory in
(3+1)D.\cite{T3141,JR7616,W8246,G8205,LM0012} Using long-range entanglements
and their string-net realization, we can obtain the simultaneous emergence of
both gauge bosons and fermions in \emph{any} dimensions and for any gauge
group.\cite{LWsta,LWstrnet,LWuni,Wqoem} This result gives us hope that maybe
every elementary particles are emergent and can be unified using local qubit
models.  Thus, long-range entanglement offer us a new option to view our world.
Maybe our vacuum is a long-range entangled state.  It is the pattern of the
long-range entanglement in the vacuum that determines the content and the
structures of observed elementary particles.  Such a picture has an
experimental prediction that is described in the next section \ref{fPred}.

We like to point out that the string-net unification of gauge bosons and
fermions is very different from the superstring theory for gauge bosons and
fermions.  In the string-net theory, gauge bosons and fermions come from the
qubits that form the space, and ``string-net'' is
simply the name that describe how qubits are organized in the ground state.  So
string-net is not a thing, but a pattern of qubits.  In the string-net theory,
the gauge bosons are waves of collective fluctuations of the string-nets, and a
fermion corresponds to one end of string.  In contrast, gauge bosons and
fermions come from strings in the superstring theory. Both gauge bosons and
fermions correspond to small pieces of strings.  Different vibrations of the
small pieces of strings give rise to different kind of particles. The fermions
in the superstring theory are put in by hand through the introduction of
Grassmann fields.}

\subsection{A falsifiable prediction of 
string-net unification of gauge interactions and Fermi statistics}
\label{fPred}

In the string-net unification of light and electrons,\cite{LWuni,LWqed} we
assume that the space is formed by a collection of qubits and the qubits form a
string-net condensed state.  Light waves are collective motions of the
string-nets, and an electron corresponds to one end of string.  Such a
string-net unification of light and electrons has a  falsifiable prediction:
\emph{all fermionic excitations must carry some gauge
charges}.\cite{LWsta,LWstrnet}

The $U(1)\times SU(2) \times SU(3)$ standard model for elementary particles
contains fermionic excitations (such as neutrons and neutrinos) that do not
carry any $U(1)\times SU(2) \times SU(3)$ gauge charge.  So according to the
string-net theory, the $U(1)\times SU(2) \times SU(3)$ standard model is
incomplete.  According to the string-net theory, our universe not only have
$U(1)\times SU(2) \times SU(3)$ gauge theory, it must also contain other gauge
theories.  Those additional  gauge theories may have a gauge group of $Z_2$ or
other discrete groups.  Those extra discrete gauge theories will lead to new
cosmic strings which will appear in very early universe.

\section{Examples of topological order: quantum liquid of unoriented strings and
emergence of statistics}

In the above, we discussed how light and electrons may emerge from a quantum
liquid of orientable strings.  We like to point out that quantum liquids of
orientable strings are not the simplest topologically ordered state.  Quantum
liquids of unoriented strings are simpler  topologically ordered states.  In
this section, we will discuss quantum liquids of unoriented strings and their
topological properties.  Using those simpler examples, we will discuss in
detail how can ends of strings become fermions, or even anyons.

\subsection{Quantum liquids of unoriented strings and the local ``dancing''
rules}

The strings in quantum liquids of unoriented strings can be realized in a
spin-1/2 model.  We can view up-spins as background and lines of down-spins as
the strings (see Fig. \ref{strspin}). Clearly, such string is unoriented.  The
simplest  topologically ordered state in such spin-1/2 system is given by the
equal-weight superposition of all closed strings:\cite{K032}
$|\Phi_{Z_2}\>=\sum_\text{all closed strings} \left |\bmm
\includegraphics[height=0.3in]{strnetS}\emm\right \> $. Such a wave function
represents a global dancing pattern that correspond to a non-trivial
topological order.

As we have mentioned before, the  global dancing pattern is determined by local
dancing rules.  What are those local rules that give rise to the global
dancing pattern $|\Phi_{Z_2}\>=\sum_\text{all closed strings} \left |\bmm
\includegraphics[height=0.3in]{strnetS}\emm\right \> $?
The first rule is that, in the
ground state, the down-spins are always connected with no open ends.
To describe the second rule,
we need to introduce
the amplitudes of close strings in the
ground state:
$\Phi\bpm
\includegraphics[height=0.3in]{strnetS}\epm$.
The ground state is given by
\begin{align}
\sum_\text{all closed strings} 
\Phi\bpm
\includegraphics[height=0.3in]{strnetS}\epm
 \left |\bmm
\includegraphics[height=0.3in]{strnetS}\emm\right \>.
\end{align}
Then the second rule relates the amplitudes of close strings in the ground state
as we change the strings locally:
\begin{align}
\label{Z2rl}
 \Phi
\bpm \includegraphics[height=0.2in]{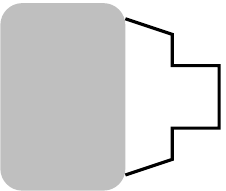} \epm  =&
\Phi
\bpm \includegraphics[height=0.2in]{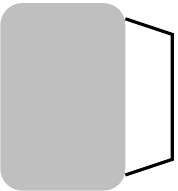} \epm ,
%\nonumber \\
&
 \Phi
\bpm \includegraphics[height=0.2in]{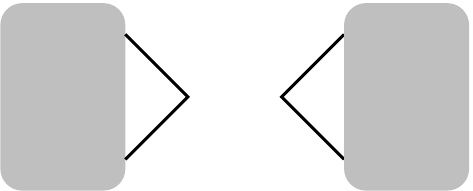} \epm  =&
\Phi
\bpm \includegraphics[height=0.2in]{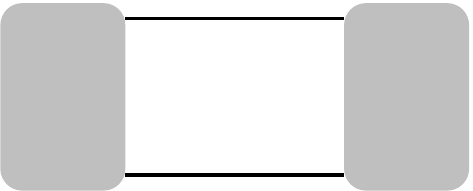} \epm,
\end{align}
In other words, if we locally deform/reconnect the strings as in Fig.
\ref{strnetSa}, the amplitude (or the ground state wave function) does not
change.

The first rule tells us that
the amplitude of a string configuration only depend on the topology
of the string configuration.
Starting from a single loop, using the local deformation and the
local reconnection in Fig. \ref{strnetSa}, we can generate all closed string
configurations with any number of loops.  So all those  closed string
configurations have the same amplitude.  Therefore, the local dancing rule
fixes the wave function to be the equal-weight superposition of all closed
strings: $|\Phi_{Z_2}\>=\sum_\text{all closed strings} \left |\bmm
\includegraphics[height=0.3in]{strnetS}\emm\right \> $.
In other words,  the local dancing rule
fixes the global dancing pattern.

If we choose another local dancing rule, then we will get a different global
dancing pattern that corresponds to a different topological order.  One of the
new choices is obtained by just modifying the sign in \eqn{Z2rl}:
\begin{align}
\label{Semrl}
 \Phi
\bpm \includegraphics[height=0.2in]{Xi1} \epm  =&
\Phi
\bpm \includegraphics[height=0.2in]{Xi} \epm ,
%\nonumber \\
&
 \Phi
\bpm \includegraphics[height=0.2in]{XijklX} \epm  =&
- \Phi
\bpm \includegraphics[height=0.2in]{XijX} \epm  .
\end{align}
We note that each local reconnection operation changes the number of loops by
1.  Thus the new local dancing rules gives rise to a wave function which has a
form $|\Phi_\text{Semi}\>=\sum_\text{all closed strings} (-)^{N_\text{loops}}
\left |\bmm \includegraphics[height=0.3in]{strnetS}\emm\right \> $, where
$N_\text{loops}$ is the number of loops.  The wave function
$|\Phi_\text{Semi}\>$ corresponds to a different global dance and a different
topological order.

\begin{figure}[tb]
\centerline{
\includegraphics[height=1.2in]{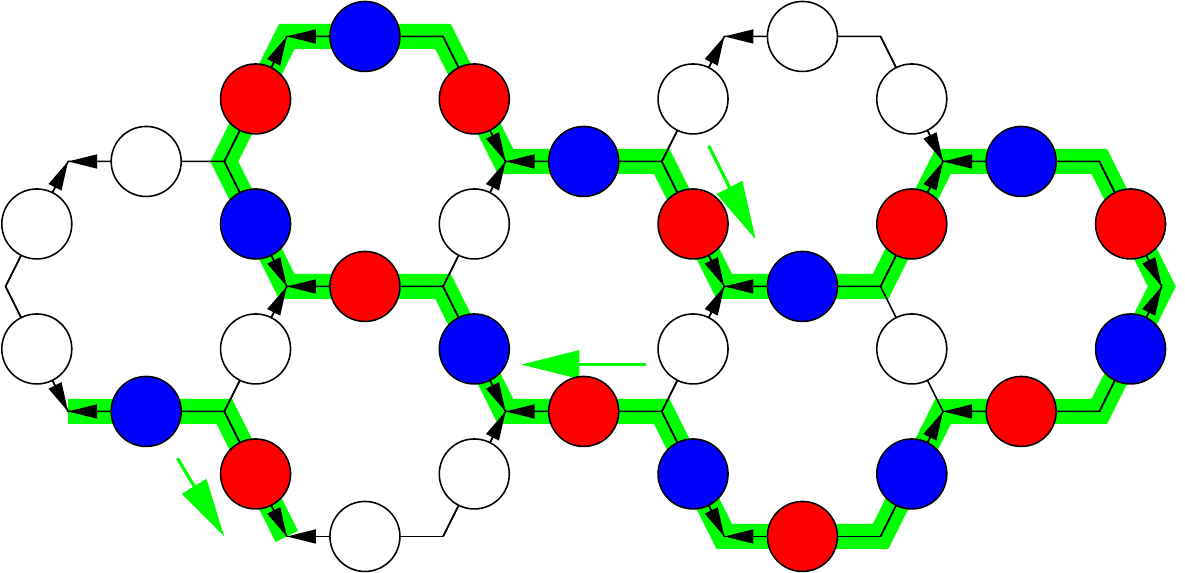}
}
\caption{
The orientable strings in a spin-1 model.  In the background of $S_z=0$ spins
(the white dots), the $S_z=1$ spins (the red dots) and the $S_z=-1$ spins (the
blue dots) form closed strings.
}
\label{strspinO}
\end{figure}

\rem{In the above, we constructed two quantum liquids of unoriented strings
in a spin-1/2 model.  Using a similar construction, we can also obtain a
quantum liquid of orientable strings which gives rise to waves satisfying
Maxwell equation as discussed before.  To obtain  quantum liquid of orientable
strings, we need to start with a spin-1 model, where spins live on the links of
honeycomb lattice (see Fig. \ref{strspinO}).  Since the honeycomb lattice is
bipartite, each link has an orientation from the A-sublattice to the
B-sublattice (see Fig. \ref{strspinO}).  The orientable strings is formed by
alternating  $S_z=\pm 1$ spins on the background of $S_z=0$ spins.  The string
orientation is given be the orientation of the links under the $S_z=1$ spins
(see Fig. \ref{strspinO}).  The superposition of the orientable strings gives
rise to quantum liquid of orientable strings.  }

\subsection{Topological properties of quantum liquids of unoriented strings}

Why the two  wave functions of unoriented strings, $|\Phi_{Z_2}\>$ and
$|\Phi_\text{Semi}\>$, have non-trivial topological orders?  This is because
the two  wave functions give rise to non-trivial topological properties.  The
two  wave functions correspond to different topological orders since they give
rise to different topological properties.  In this section, we will discuss two
topological properties: emergence of fractional statistics and topological
degeneracy on compact spaces.

\subsubsection{Emergence of Fermi and fractional statistics}

The two topological states in two dimensions contain only closed strings, which
represent the ground states.  If the wave functions contain open strings (\ie
have non-zero amplitudes for open string states), then the ends of the open
strings will correspond to point-like topological excitations above the ground
states.  Although an open string is an extended object, its middle part merge
with the strings already in the ground states and is unobservable.  Only its
two ends carry energies and correspond to two point-like particles.  

We note that such a point-like particle from an end of string cannot be created
alone.  Thus an end of string correspond to a topological point defect, which
may carry fractional quantum numbers.  This is because an open string as a
whole always carry non-fractionalized quantum numbers.  But an open string
corresponds to \emph{two} topological point defects from the two ends.  So we
cannot say that each end of string carries non-fractionalized quantum numbers.
Some times, they do carry fractionalized quantum numbers.

Let us first consider the defects in the $|\Phi_{Z_2}\>$ state.  To understand
the fractionalization, let us first consider the spin of such a defect to see if the
spin is fractionalized or not.\cite{FFN0683,Wang10}  An end of string can be
represented by 
\begin{align}
\big |\bmm \includegraphics[scale=0.33]{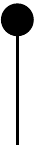}\emm \big \>_\text{def} 
= 
\big |\bmm \includegraphics[scale=0.33]{def1}\emm \big \>+
\big |\bmm \includegraphics[scale=0.33]{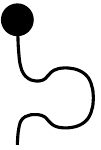}\emm \big \>+
\big |\bmm \includegraphics[scale=0.33]{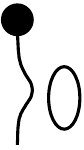}\emm \big \>+ ...
.
\end{align}
which is an equal-weight superposition of all
string states obtained from the
deformations and the reconnections of $\bmm \includegraphics[scale=0.33]{def1}\emm$.

Under a $360^\circ$ rotation, the end of string is changed to $\big |\bmm
\includegraphics[scale=0.33]{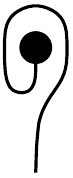}\emm \big \>_\text{def} $, which is an equal
weight superposition of all string states obtained from the
deformations and the reconnections of $\bmm \includegraphics[scale=0.33]{def3}\emm$.
Since $\big |\bmm \includegraphics[scale=0.33]{def1}\emm \big \>_\text{def} $
and $\big |\bmm \includegraphics[scale=0.33]{def3}\emm \big \>_\text{def} $ are
alway different, $\big |\bmm \includegraphics[scale=0.33]{def1}\emm \big
\>_\text{def} $ is not an eigenstate of $360^\circ$ rotation and does not carry
a definite spin.

To construct the  eigenstates of $360^\circ$ rotation, let us make a
$360^\circ$ rotation to $\big |\bmm \includegraphics[scale=0.33]{def3}\emm \big
\>_\text{def}$.  To do that, we first use the string reconnection move in Fig.
\ref{strnetSa}, to show that $\big |\bmm \includegraphics[scale=0.33]{def3}\emm
\big \>_\text{def} = \big |\bmm \includegraphics[scale=0.33]{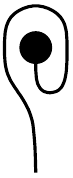}\emm \big \>_\text{def} $.  A
$360^\circ$ rotation on $\big |\bmm \includegraphics[scale=0.33]{def2}\emm \big
\>_\text{def} $ gives us $\big |\bmm \includegraphics[scale=0.33]{def1}\emm \big \>_\text{def} $.

We see that the $360^\circ$ rotation
exchanges $\big |\bmm \includegraphics[scale=0.33]{def1}\emm \big \>_\text{def} $
and $\big |\bmm \includegraphics[scale=0.33]{def3}\emm \big \>_\text{def} $.
Thus the  eigenstates of
$360^\circ$ rotation are given by
$\big |\bmm \includegraphics[scale=0.33]{def1}\emm \big \>_\text{def} + \big |\bmm
\includegraphics[scale=0.33]{def3}\emm \big \>_\text{def} $ with
eigenvalue 1, and by $\big |\bmm
\includegraphics[scale=0.33]{def1}\emm \big \>_\text{def} - \big |\bmm
\includegraphics[scale=0.33]{def3}\emm \big \>_\text{def} $ with eigenvalue $-1$.
So the particle $\big |\bmm \includegraphics[scale=0.33]{def1}\emm \big \>_\text{def} +
\big |\bmm \includegraphics[scale=0.33]{def3}\emm \big \>_\text{def} $ has a spin 0 (mod
1), and the particle $\big |\bmm \includegraphics[scale=0.33]{def1}\emm \big \>_\text{def}
- \big |\bmm \includegraphics[scale=0.33]{def3}\emm \big \>_\text{def} $ has a spin 1/2
(mod 1).

\begin{figure}[tb]
\centerline{
\includegraphics[height=1.2in]{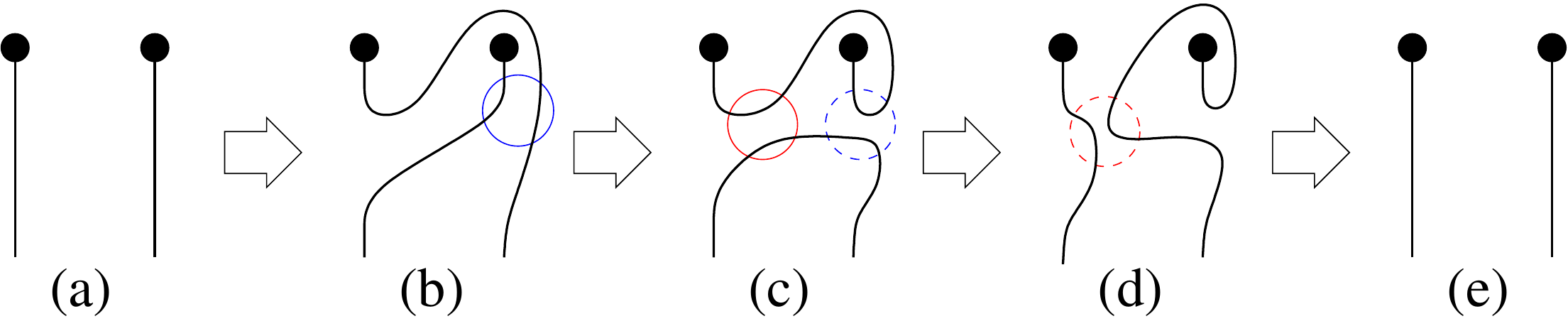}
}
\caption{
Deformation of strings and two reconnection moves, plus an exchange of two ends
of strings and a $360^\circ$ rotation of one of the end of string, change the
configuration (a) back to itself.  Note that from (a) to (b) we exchange the
two ends of strings, and from (d) to (e) we rotate of one of the end of string
by $360^\circ$.
The combination of those moves do not generate any phase.
}
\label{exch}
\end{figure}

%\begin{figure}[tb]
%\centerline{
%\includegraphics[height=1.2in]{exch1}
%}
%\caption{
%Deformation of strings and two reconnection moves generate a motion of moving
%one end of string around the other end of string and a $360^\circ$ rotation of
%the both of ends of strings.
%Note that from (a) to (b) we move 
%one end of string
%around the other end of string.
%}
%\label{exch1}
%\end{figure}

If one believes in the spin-statistics theorem, one may guess that the particle
$\big |\bmm \includegraphics[scale=0.33]{def1}\emm \big \>_\text{def} + \big |\bmm
\includegraphics[scale=0.33]{def3}\emm \big \>_\text{def} $ is a boson and the particle
$\big |\bmm \includegraphics[scale=0.33]{def1}\emm \big \>_\text{def} - \big |\bmm
\includegraphics[scale=0.33]{def3}\emm \big \>_\text{def} $ is a fermion.  This guess is
indeed correct.  Form Fig. \ref{exch}, we see that we can use deformation of
strings and two reconnection moves to generate an exchange of two ends of
strings and a $360^\circ$ rotation of one of the end of string.  Such
operations allow us to show that Fig. \ref{exch}a and  Fig. \ref{exch}e have
the same amplitude, which means that an exchange of two ends of strings
followed by a $360^\circ$ rotation of one of the end of string do not generate
any phase.  This is nothing but the spin-statistics theorem.  

%In Fig. \ref{exch1}, we also see
%that the deformation of strings and two reconnection moves generate a motion of
%moving one end of string around the other end of string and a $360^\circ$
%rotation to the both of ends of strings.  Such moves do not generate any
%phases.  Thus moving one end of string around the other end of string generate
%the same phase as doing  a $360^\circ$ rotation to the both of ends of strings.
%Thus moving the particle
%$\big |\bmm \includegraphics[scale=0.33]{def1}\emm \big \>_\text{def} + \big |\bmm
%\includegraphics[scale=0.33]{def3}\emm \big \>_\text{def} $ around the particle
%$\big |\bmm \includegraphics[scale=0.33]{def1}\emm \big \>_\text{def} - \big |\bmm
%\includegraphics[scale=0.33]{def3}\emm \big \>_\text{def} $
%generate a phase $-1$.
%\ie the two particles have a mutual semion statistics.

The emergence of Fermi statistics in the $|\Phi_{Z_2}\>$
state of a purely bosonic spin-1/2 model
indicates that the state is a topologically ordered state.  We also see
that the $|\Phi_{Z_2}\>$ state has a bosonic quasi-particle $\big |\bmm
\includegraphics[scale=0.33]{def1}\emm \big \>_\text{def} + \big |\bmm
\includegraphics[scale=0.33]{def3}\emm \big \>_\text{def} $, and a fermionic quasi-particle
$\big |\bmm \includegraphics[scale=0.33]{def1}\emm \big \>_\text{def} - \big |\bmm
\includegraphics[scale=0.33]{def3}\emm \big \>_\text{def} $.  The bound state of the above
two particles is a boson (not a fermion) due to their  mutual semion
statistics.  Such quasi-particle content agrees exactly with the $Z_2$ gauge
theory which also has three type of non-trivial quasiparticles excitations, two
bosons and one fermion.  In fact, the low energy effective theory of the
topologically ordered state $|\Phi_{Z_2}\>$ is the $Z_2$ gauge theory and we
will call  $|\Phi_{Z_2}\>$ a $Z_2$ topologically ordered state.

Next, let us consider the defects in the $|\Phi_\text{Semi}\>$ state.  
Now
\begin{align}
\big |\bmm \includegraphics[scale=0.33]{def1}\emm \big \>_\text{def} 
= 
\big |\bmm \includegraphics[scale=0.33]{def1}\emm \big \>+
\big |\bmm \includegraphics[scale=0.33]{def1a}\emm \big \>-
\big |\bmm \includegraphics[scale=0.33]{def1b}\emm \big \>+ ...
.
\end{align}
and a similar expression for $\big |\bmm \includegraphics[scale=0.33]{def3}\emm
\big \>_\text{def}$, due to a
change of the local dancing rule
for reconnecting the strings (see \eqn{Semrl}).  
Using the string reconnection move in Fig.
\ref{strnetSa}, we find that $\big |\bmm \includegraphics[scale=0.33]{def3}\emm
\big \>_\text{def} = - \big |\bmm \includegraphics[scale=0.33]{def2}\emm \big
\>_\text{def} $.  So a $360^\circ$ rotation, changes $(\big |\bmm
\includegraphics[scale=0.33]{def1}\emm \big \>_\text{def}, \big |\bmm
\includegraphics[scale=0.33]{def3}\emm \big \>_\text{def} )$ to $( \big |\bmm
\includegraphics[scale=0.33]{def3}\emm \big \>_\text{def}, -\big |\bmm
\includegraphics[scale=0.33]{def1}\emm \big \>_\text{def} )$.  We find that
$\big |\bmm \includegraphics[scale=0.33]{def1}\emm \big \>_\text{def} + \imth
\big |\bmm \includegraphics[scale=0.33]{def3}\emm \big \>_\text{def} $ is the
eigenstate of the $360^\circ$ rotation with eigenvalue $-\imth$, and $\big
|\bmm \includegraphics[scale=0.33]{def1}\emm \big \>_\text{def} - \imth \big
|\bmm \includegraphics[scale=0.33]{def3}\emm \big \>_\text{def} $ is the other
eigenstate of the $360^\circ$ rotation with eigenvalue $\imth$.  So the
particle $\big |\bmm \includegraphics[scale=0.33]{def1}\emm \big \>_\text{def}
+ \imth \big |\bmm \includegraphics[scale=0.33]{def3}\emm \big \>_\text{def} $
has a spin $-1/4$, and the particle $\big |\bmm
\includegraphics[scale=0.33]{def1}\emm \big \>_\text{def} - \imth \big |\bmm
\includegraphics[scale=0.33]{def3}\emm \big \>_\text{def} $ has a spin $1/4$.
The spin-statistics theorem is still valid for $|\Phi_\text{Semi}\>_\text{def}$
state, as one can see form Fig. \ref{exch}.  So, the particle $\big |\bmm
\includegraphics[scale=0.33]{def1}\emm \big \>_\text{def} + \imth\big |\bmm
\includegraphics[scale=0.33]{def3}\emm \big \>_\text{def} $ and particle $\big
|\bmm \includegraphics[scale=0.33]{def1}\emm \big \>_\text{def} - \imth\big
|\bmm \includegraphics[scale=0.33]{def3}\emm \big \>_\text{def} $ have
fractional statistics with statistical angles of semion: $\pm \pi/2$.  Thus the
$|\Phi_\text{Semi}\>$ state contains a non-trivial topological order.  We will
call such a topological order a double-semion topological order.

It is amazing to see that the long range quantum entanglements in string liquid
can give rise to fractional spin and fractional statistics, even from a purely
bosonic model.  Fractional spin and Fermi statistics are two of most mysterious
phenomena in natural.  Now, we can understand them as merely a phenomenon of
long-range quantum entanglements.  They are no longer mysterious.

\subsubsection{Topological degeneracy}

The $Z_2$ and the  double-semion topological states (as well as many other
topological states) have another important topological property: topological
degeneracy.\cite{Wtop,Wrig}  Topological degeneracy is the ground state
degeneracy of a gapped many-body system that is robust against any local
perturbations as long as the system size is large.

Topological degeneracy can be used as protected qubits which allows us to
perform topological quantum computation.\cite{K032} It is believed that the
appearance of topological degeneracy implies the topological order (or
long-range entanglements) in the ground state.\cite{Wtop,Wrig} Many-body states
with topological degeneracy are described by topological quantum field theory
at low energies.\cite{W8951}

The simplest topological  degeneracy appears when we put topologically ordered
states on compact spaces with no boundary.  We can use the global dancing
pattern to understand the  topological  degeneracy.  We know that the local
dancing rules determine the global dancing pattern.  On a sphere, the  local
dancing rules determine a unique global dancing pattern.  So the ground state
is non-degenerate.  However on other  compact spaces, there can be several
global dancing patterns that all satisfy the the  local dancing rules. In this
case, the ground state is degenerate.

\begin{figure}[tb]
\centerline{
\includegraphics[height=1.4in]{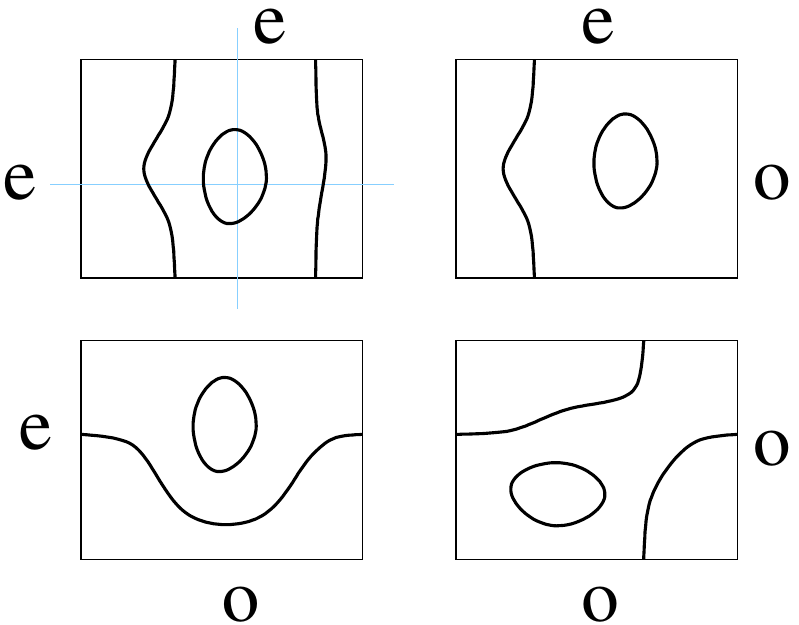}
}
\caption{
 On a torus, the closed string configurations
can be divided into four sectors, depending on even or
odd number of strings crossing the x- or y-axises.
}
\label{z2eo}
\end{figure}

For the $Z_2$ topological state on torus, the local dancing rule relate the
amplitudes of the string configurations that differ by a string reconnection
operation in Fig. \ref{strnetSa}.  On a torus, the closed string configurations
can be divided into four sectors (see Fig. \ref{z2eo}), depending on even or
odd number of strings crossing the x- or y-axises.  The string reconnection
move only connect the string configurations among each sector.  So the
superposition of the string configurations in each sector represents a
different global dancing pattern and a different degenerate ground state.
Therefore, the local dancing rule for the  $Z_2$ topological order gives rise
to four fold degenerate ground state on torus.\cite{Wsrvb}

Similarly,  the double-semion topological order also gives rise to four fold
degenerate ground state on torus.

%\section{Examples of topological order: FQH states}

\section{A macroscopic definition and the characterization of topological order}

So far in this paper, we
discussed topological order using an intuitive dancing picture.  Then we
discussed a few simple examples.  In the rest of this paper, we will give a
more rigorous description and a systematic understanding of topological order
and its essence.\cite{Wtop,Wrig} Historically, the  more rigorous description
of topological order was obtained before the  intuitive dancing picture and the
simple examples of  topological order discussed in the previous part of the
paper.

\begin{figure}[tb]
\centerline{
\includegraphics[scale=0.6]{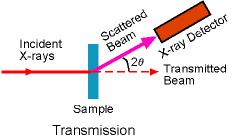} 
\hfil
\includegraphics[scale=0.3]{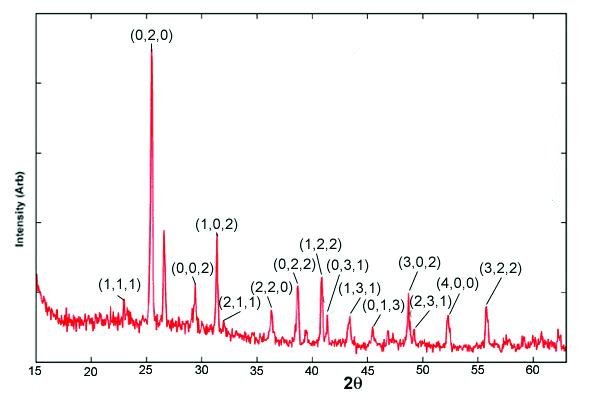}
}
\caption{
A X-ray diffraction pattern defines/probes the crystal order.
}
\label{xray}
\end{figure}

First, we would like to give a physical definition of topological order
(at least in 2+1 dimensions).  Here, we like to point out that \emph{to define
a physical concept is to design experiments or numerical calculations that
allow us probe and characterize the  concept}.  For example, the concept of
superfluid order, is defined by zero viscosity and the quantization of
vorticity, and the concept of crystal order is defined by X-ray diffraction
experiment (see Fig.  \ref{xray}).

\begin{table}[tb]
 \centering
 \begin{tabular}{ |c|c|}
 \hline
\textbf{Order} & \textbf{Experiment} \\
 \hline
Crystal order & X-ray diffraction\\
 \hline
Ferromagnetic order & Magnetization\\
 \hline
Anti-ferromagnetic order & Neutron scattering\\
 \hline
Superfluid order & Zero-viscosity \& vorticity quantization\\
 \hline
 \hline
Topological order & Topological degeneracy, \\
(Global dancing pattern) & non-Abelian geometric phase \\
 \hline
 \end{tabular}
\caption{
Symmetry breaking orders can be probed/defined through linear responses. 
But topological order cannot be
probed/defined through  linear responses.
We need topological probes to define topological orders.
}
\label{tab1}
\end{table}

\begin{figure}[tb]
\centerline{
\includegraphics[scale=0.4]{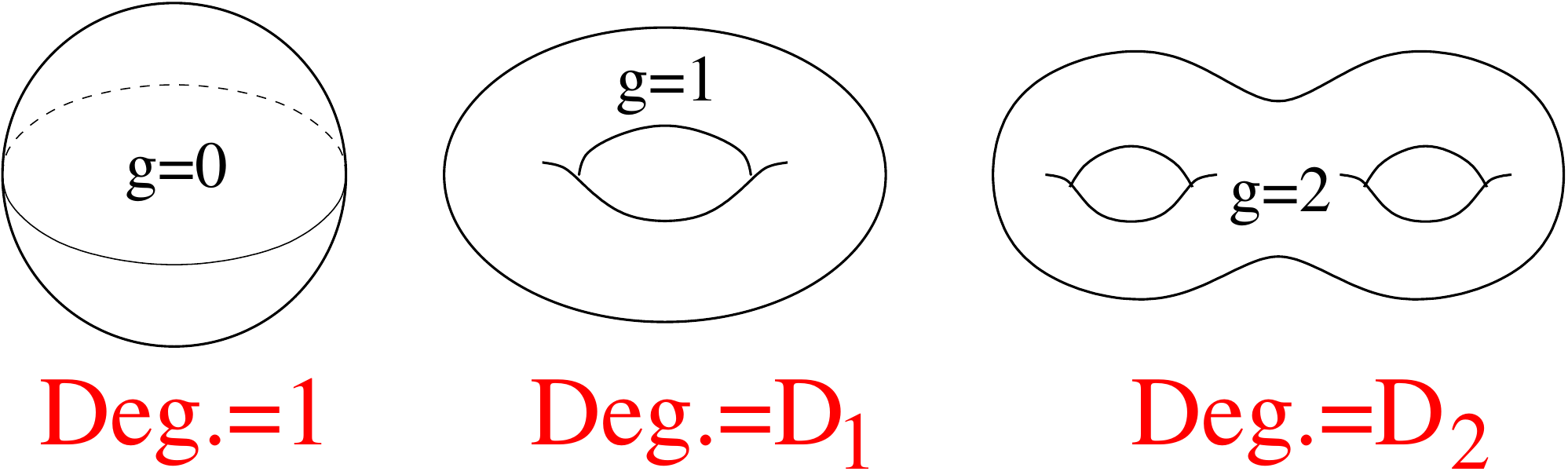}
}
\caption{The topological ground state degeneracies of topologically ordered
states depend on the topology of the space, such as the genus $g$ of two
dimensional closed surfaces.}
\label{g0g1g2}
\end{figure}

The experiments that we use to define/characterize superfluid order and crystal
order are linear responses.  Linear responses are easily accessible in
experiments and the symmetry breaking order that they define are easy to
understand (see Table \ref{tab1}).  However, topological order is such a new and elusive order that it
cannot be probed/defined by any linear responses.  To probe/define topological
order we need to use very unusual ``topological'' probes.
%This may be the reason why the concept of topological order
%was only introduced 7 year after the discovery of FQH states:
In 1989, we conjectured that topological order
can be completely defined/characterized by using only two topological
properties (at least in 2+1 dimensions):\cite{Wrig}\\
(1) Topological ground state degeneracies on closed spaces 
of various topologies.  (see Fig. \ref{g0g1g2}).\cite{Wtop}\\
(2) Non-Abelian geometric phases\cite{WZ8411} of those degenerate ground states
from deforming the spaces (see Fig. \ref{modtrns}).\cite{Wrig,KW9327}.\\
It was through such topological probes that allowed us to introduce the concept
of topological order.  \emph{Just like zero viscosity and the quantization of
vorticity define the concept of superfluid order, the topological degeneracy
and the non-Abelian geometric phases of the degenerate ground states define the
concept of topological order.}

\subsection{What is ``topological ground state degeneracy''}

Topological ground state degeneracy, or simply, topological degeneracy is a
phenomenon of quantum many-body systems, that the ground state of a gapped
many-body system become degenerate in the large system size limit, and that
such a degeneracy \emph{cannot be lifted by any local perturbations} as long as
the system size is large.\cite{Wtop,WNtop,WZ9817,HWcnt}
%
%For a gapped many-body system of a large finite size, its energy spectrum at
%low energies is shown in Fig. \ref{gap}.  The states below the energy gap are
%the nearly degenerate ground states.  The number of the  nearly degenerate
%%ground states is the  ground state degeneracy.  If the ground state degeneracy
%is robust against any local perturbations in large system size limit, then the
%ground state degeneracy becomes topological
%degeneracy.\cite{Wtop,WNtop,WZ9817}
The topological degeneracy for a given system usually is different for
different topologies of space.\cite{HR8529}  For example, for the $Z_2$
topologically ordered state in two dimensions,\cite{WWZ8913} the topological
degeneracy is $D_g=4^g$ on genus $g$ Riemann surface (see Fig. \ref{g0g1g2}).

People usually attribute the ground state degeneracy to symmetry.  But
topological degeneracy, being robust against any local perturbations, is not
due to symmetry. So the very existence of topological degeneracy is a
surprising and amazing phenomenon.  Such an amazing phenomenon defines the
notion of topological order.  As a comparison, we know that the existence of
zero-viscosity is also an amazing phenomenon, and such an amazing phenomenon
defines the notion of superfluid order.  So topological degeneracy,
%in topological order, 
playing the role of zero-viscosity in superfluid order,
implies the existence of a new kind of quantum phase -- topologically ordered
phases.

\subsection{What is ``non-Abelian geometric phase of topologically degenerate
states''}

\begin{figure}[tb]
\centerline{
\includegraphics[scale=0.6]{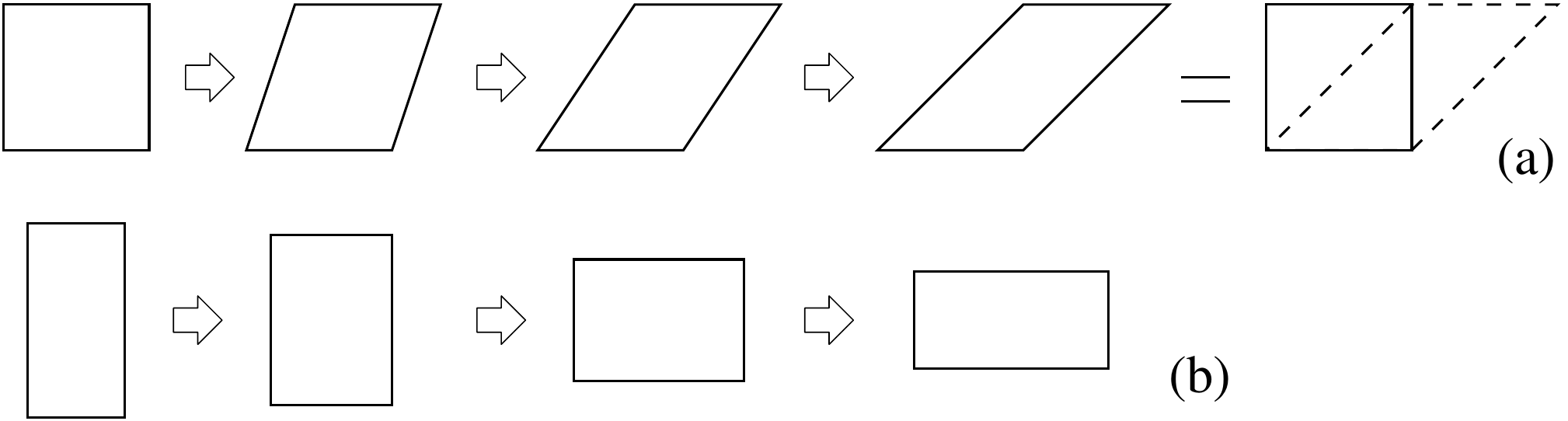}
}
\caption{
(a) The shear deformation of a torus generate a (projective) non-Abelian
geometric phase $T$, which is a generator of
a projective representation modular transformation.
The last shear-deformed torus is the same as the original
torus after a coordinate transformation:
$x\to x+y$, $y\to y$.
(b) The squeezing deformation of a torus generate a (projective) non-Abelian
geometric phase $S$, which is the other generator of
a projective representation modular transformation.
The last squeeze-deformed torus is the same as the original
torus after a coordinate transformation:
$x\to y$, $y\to -x$.
}
\label{modtrns}
\end{figure}

%\begin{figure}[tb]
%\centerline{
%\includegraphics[scale=0.6]{gap}
%}
%\caption{
%The low energy spectrum of a gapped many-body system.  The energy splitting
%$\veps$ of nearly degenerate ground states approaches to zero in large system
%size limit, while the energy gap $\Del$ above the ground states approach to a
%constant value.
%}
%\label{gap}
%\end{figure}

\rem{
However, the ground state degeneracy is not enough to completely
characterize/define topological order.  Two different topological orders may
have exactly the same topological degeneracy on space of any topology.  We
would like to find, as many as possible, quantum numbers associated with the
degenerate ground states, so that by measuring these quantum numbers we can
completely characterize/define topological order.  The non-Abelian geometric
phases  of topologically degenerate states are such quantum
numbers.\cite{Wrig,KW9327}

The non-Abelian geometric phase is a unitary matrix $U$ that can be calculated
from an one parameter family of gapped Hamiltonians $H_g$, $g\in [0,1]$,
provided that $H_0=H_1$.\cite{WZ8411} $U$ is a one by one matrix if there is
only one ground state below the gap. $U$ is $n$ dimensional if the ground state
degeneracy is $n$ for all $g\in [0,1]$.  
%(Note that the statement `` the ground state degeneracy is $n$'' implies that
%the energy gap above the $n$ ground states are finite in large-system-size
%limit.)

To use non-Abelian geometric phases to characterize/define topological order,
let us put the many-body state on a torus,\cite{Wrig,KW9327,ZGT1251,ZV1224} and
perform a ``shear'' deformation of the torus to obtain an one parameter family
of gapped Hamiltonians that form a loop (\ie $H_0=H_1$) (see Fig.
\ref{modtrns}a).  The  non-Abelian geometric phase obtained this way is denoted
as $T$.  Similarly, a ``squeezing'' deformation of the torus gives rise to
another non-Abelian geometric phase $S$.  Both $S$ and $T$ are $D_1$
dimensional unitary matrices where $D_1$ is the topological degeneracy on
torus.  For different deformation paths that realize the loops in Fig.
\ref{modtrns}, $S$ and $T$ may be different.  However, because the ground state
degeneracy is robust, the difference is \emph{only} in the total phase factors.
Since the two deformations in Fig.  \ref{modtrns} generate the modular
transformations, thus $S$ and $T$ generate a projective representation of the
modular transformations.  It was conjectured that $S$ and $T$ (or the
projective representation of the  modular transformations) provides a complete
characterization and definition of topological orders in 2+1
dimensions.\cite{Wrig,KW9327} }

%\subsubsection{Topological properties define topological order}

%To summarize, topological order is a new kind of order in zero-temperature
%phase of matter (also known as quantum matter).  Topological order is beyond
%the Landau symmetry-breaking description. It cannot be described by local order
%parameters and long range correlations.  However, topological orders can be
%described/defined by a new set of quantum phenomena: topological ground state
%degeneracy and the non-Abelian geometric phases of those degenerate ground
%states.
%quasi-particle fractional statistics, edge states, topological entanglement
%entropy, etc.

%We know that the (experimental) discoveries of zero-viscosity and the
%quantization of vorticity lead to discovery of superfluid order.  The different
%superfluid orders are classified by the different ways in which the  vorticity
%is quantized.  Similarly, the (theoretical) discoveries of topological
%degeneracy and the non-Abelian geometric phases of those degenerate ground
%states lead to the discovery of a new order -- topological order.
%
%In 2+1 dimensions, the  non-Abelian geometric phases on torus generate a
%projective representation of modular transformation.  So we may use the
%different projective representations of modular transformation as a starting
%point to classify different topological orders in 2+1 dimensions.  It would be
%really interesting to develop a macroscopic theory of topological order based
%on the   projective representations of modular transformation.

\subsection{The essence of topological orders}

\rem{
C. N. Yang once asked: the microscopic theory of fermionic superfluid and
superconductor, BCS theory, capture the essence of the  superfluid and
superconductor, but what is this essence?  This question led him to develop the
theory of off-diagonal long range order\cite{Y6294} which reveal the essence of
superfluid and superconductor. In fact long range order is the essence of any
symmetry breaking order.  

Similarly, we may ask: Laughlin's theory for FQH effect
capture the essence of the FQH effect, but what is this essence?  Our answer is
that the topological order 
%defined in the last section
defined by the topological ground state degeneracy
and the non-Abelian geometric phases of those degenerate ground states 
is the essence of FQH effect.

One may disagree with the above statement by pointing out that the  essence of
FQH effect should be the quantized Hall conductance.  However, such an opinion
is not quite correct, since even after we break the particle number
conservation (which breaks the quantized Hall conductance), a FQH state is
still a non-trivial state with  a quantized thermal Hall
conductance.\cite{KF9732}  The non-trivialness of FQH state does not rely on
any symmetry (except the conservation of energy).  In fact, the topological
degeneracy and the non-Abelian geometric phases discussed above are the essence
of FQH states which can be defined even without any symmetry.  They provide a
characterization and definition of topological order that does not rely on any
symmetry.  We would like to point out that the topological entanglement entropy
is another way to characterize the topological order  without any
symmetry.\cite{KP0604,LWtopent}

}

%\subsection{Comments on other characterizations of topological order}
%
%There are many other macroscopic probes/characterizations of topological order,
%that does not rely on any symmetry.  In this subsection we will briefly discuss
%some of them.
%
%\subsubsection{Quantum entanglement}
%
%One can use quantum entanglement to probe and characterize topological
%order. Such a characterization is very important conceptually.  It points out
%an important connection between topological order and long-range  quantum
%entanglements.
%
%

%Although topologically ordered states usually appear in strongly interacting
%boson/fermion systems, a simple kind of topological order can also appear in
%free fermion systems. This kind of topological order correspond in integral
%quantum Hall state, which can be characterized by the Chern number of the
%filled energy band if we consider integral quantum Hall state on a lattice.
%Theoretical calculations have proposed that such Chern number can be measured
%for a free fermion system experimentally. 
%
%It is also well known that such a Chern number can be measured (may be
%indirectly) by edge states.

\section{The microscopic description of topological order}
 
After the experimental discovery of superconducting order via zero-resistance
and Meissner effect,\cite{O1122} it took 40  years to obtain the microscopic
understanding of superconducting order through the condensation of fermion
pairs.\cite{BCS5775}  However, we are luckier for topological orders. After the
theoretical  discovery of topological order via the topological degeneracy and
the non-Abelian geometric phases of the degenerate ground states,\cite{Wrig} it
took only 20 years to obtain the microscopic understanding of topological
order: topological order is due to long-range entanglements and topological
order is simply pattern of long-range entanglements.\cite{CGW1038}  In this
section, we will explain such a microscopic understanding.

\subsection{Local unitary transformations}

\label{LUt}

The long-range entanglements is defined through local unitary (LU)
transformations.  LU transformation is an important concept which is directly
related to the definition of quantum phases.\cite{CGW1038}  In this section, we
will give a short review of LU
transformation.\cite{LWstrnet,VCL0501,V0705,CGW1038}

Let us first introduce local unitary evolution.  A LU evolution is
defined as the following unitary operator that act on the degrees of freedom in
a quantum system:
\begin{align}
\label{LUdef}
  \cT[e^{-i\int_0^1 dg\, \t H(g)}]
\end{align}
where $\cT$ is the path-ordering operator and $\t H(g)=\sum_{\v i} O_{\v i}(g)$
is a sum of local Hermitian operators.  Two gapped quantum states belong to the
same phase if and only if they are related by a LU
evolution.\cite{HWcnt,BHM1044,CGW1038}

\begin{figure}[tb]
\begin{center}
\includegraphics[scale=0.5]{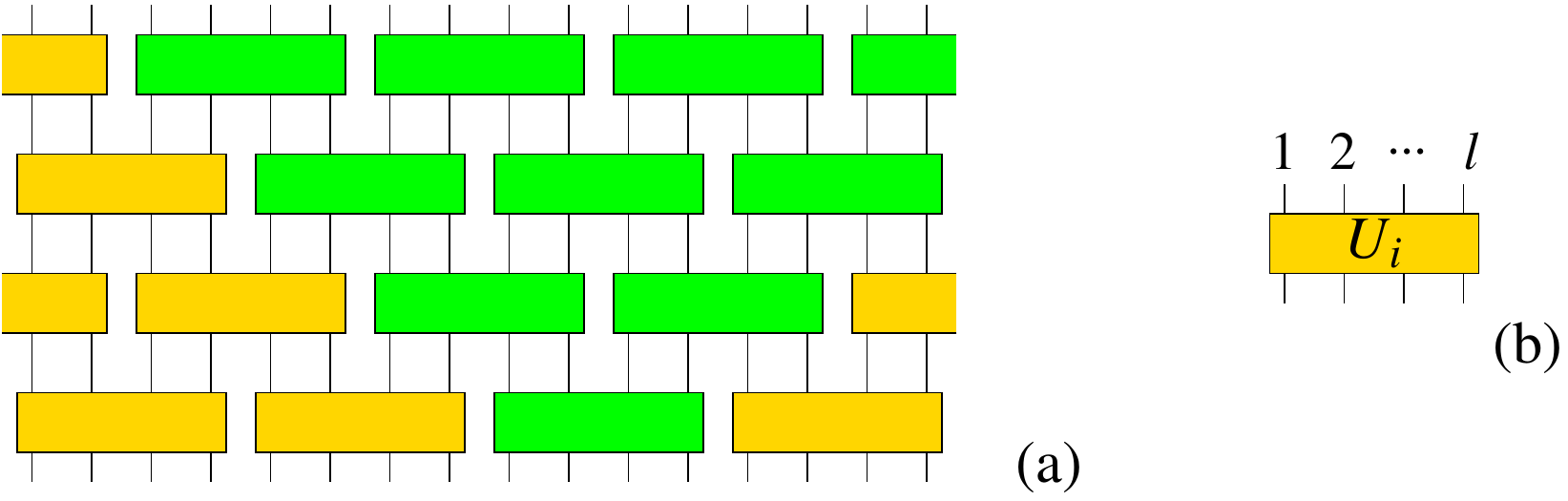}
%Fig. 3
\end{center}
\caption{
(a) A graphic representation of a quantum circuit, which is form by (b) unitary
operations on blocks of finite size $l$. The green shading represents a causal
structure.
}
\label{qc}
\end{figure}

The LU evolutions is closely related to \emph{quantum circuits with finite
depth}.  To define quantum circuits, let us introduce  piecewise local unitary
operators.  A piecewise local unitary operator has a form
\begin{equation*}
 U_{pwl}= \prod_{i} U^i
\end{equation*}
where $\{ U^i \}$ is a set of unitary operators that act on non overlapping
regions. The size of each region is less than some finite number $l$. The
unitary operator $U_{pwl}$ defined in this way is called a piecewise local
unitary operator with range $l$.  A quantum circuit with depth $M$ is given by
the product of $M$ piecewise local unitary operators:
\begin{equation*}
 U^M_{circ}= U_{pwl}^{(1)} U_{pwl}^{(2)} \cdots U_{pwl}^{(M)}
\end{equation*}
We will call $U^M_{circ}$ a LU transformation.  In quantum information theory,
it is known that finite time unitary evolution with local Hamiltonian (LU
evolution defined above) can be simulated with constant depth quantum circuit
(\ie a  LU transformation) and vice-verse:
\begin{align}
  \cT[e^{-i\int_0^1 dg\, \t H(g)}] =U^M_{circ}.
\end{align}
So two gapped quantum states belong to the same phase if and only if they are
related by a LU transformation.

%In this proposal, we will use the LU transformations to simplify gapped quantum
%states within the same phase.  This allows us to gain a deeper understanding and
%even to classify gapped quantum phases.

\subsection{Topological orders and long-range entanglements}

\begin{figure}[b]
\begin{center}
\includegraphics[scale=0.6]{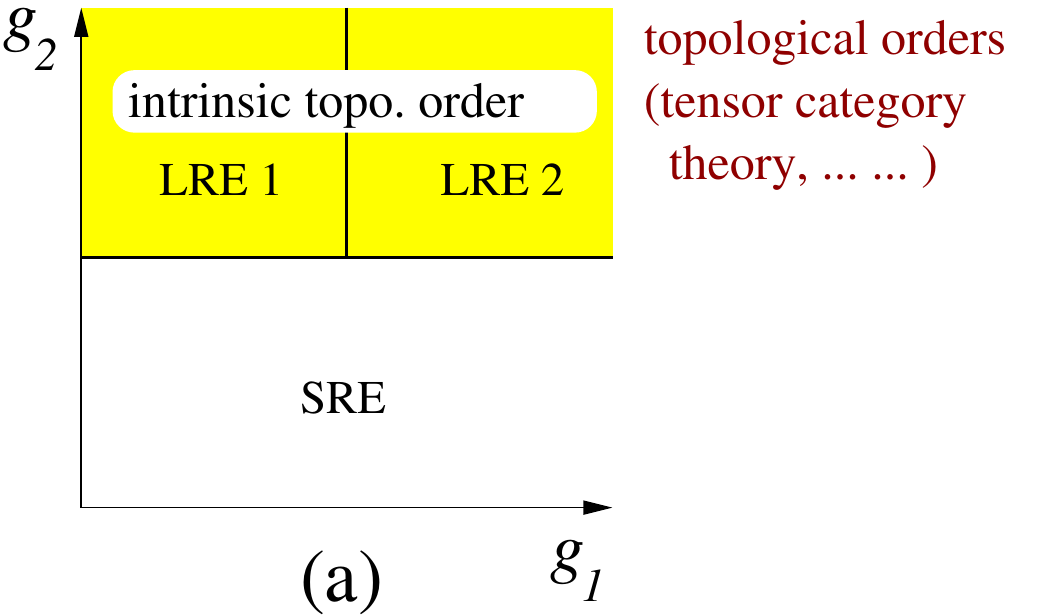}
%Fig. 1
\end{center}
\caption{
The possible gapped phases for a class of Hamiltonians $H(g_1,g_2)$ without any
symmetry restriction.  Each phase is labeled by its entanglement properties.
SRE stands for short range entanglement, and LRE for long range entanglement
which correspond to topologically ordered phases. 
}
\label{topphs}
\end{figure}

The notion of LU transformations leads to the
following more general and more systematic picture of phases and phase
transitions (see Fig.  \ref{topphs}).\cite{CGW1038} For gapped quantum systems
without any symmetry, their quantum phases can be divided into two classes:
short range entangled (SRE) states and long range entangled (LRE) states.

SRE states are states that can be transformed into direct product states via LU
transformations. All SRE states can be transformed into each other via  LU
transformations. So all SRE states belong to the same phase (see Fig.
\ref{topphs}a).

LRE states are states that cannot be transformed into  direct product states
via LU transformations.  It turns out that, many LRE states also cannot be
transformed into each other. The LRE states that are not connected via LU
transformations belong to different classes and represent different quantum
phases.  Those different quantum phases are nothing but the topologically
ordered phases.  So, topological order is pattern of long-range
entanglements.  

Such a understanding of topological order in terms of long-range entanglements
lead to a systematic description of boundary-gapped (BG) topological orders in
2+1 dimensions,\cite{LWstrnet,H0904,CGW1038,GWW1017} in terms of spherical
fusion category.\cite{Wang10}  (Here, an BG topological order is a long-range
entangled phase which can have an gapped edge or gapped entanglement
spectrum.\cite{LH0804})

In (2+1)D, BG topological orders can be viewed as string-net liquids, where the
global dancing patterns (\ie topological orders or patterns of long-range
entanglements) can be determined by local dancing rules that are similar to
\eqn{Z2rl} and \eqn{Semrl}.  For those more general  BG topological orders, the
strings in the string-net liquid may have several types labeled by
$i,j,...=0,1,...,N$, and they may join to form a string-net.  The local dancing
rules relate the amplitudes of string-net configurations that only differ by
small local transformations. To write down a set of 
local rules, one first chooses a real tensor $d_i$ and an complex
tensor $F^{ijm}_{kln}$
%, $\om^k_{ij}$ 
where the indices $i,j,k,l,m,n$ run over the different string types $0,1,...N$.
The local dancing rules are then given by:
\begin{align}
\label{lclrl}
 \Phi
\bpm \includegraphics[height=0.3in]{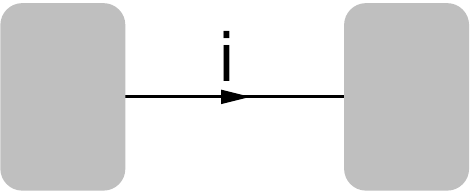} \epm  &=
\Phi 
\bpm \includegraphics[height=0.3in]{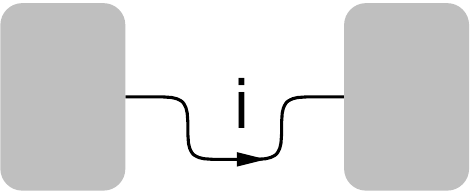} \epm
\nonumber\\
 \Phi
\bpm \includegraphics[scale=0.35]{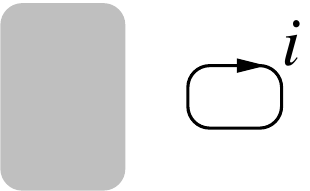} \epm  &=
d_i\Phi 
\bpm \includegraphics[scale=0.35]{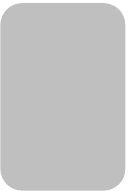} \epm
\nonumber\\
\Phi
\bpm \includegraphics[scale=0.35]{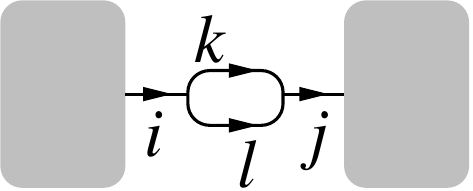} \epm  &=
\delta_{ij}
\Phi 
\bpm \includegraphics[scale=0.35]{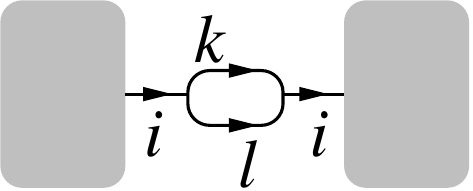} \epm
\nonumber\\
\Phi
\bpm \includegraphics[scale=0.35]{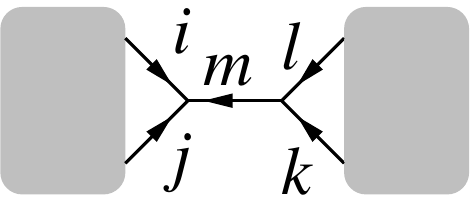} \epm  &=
\sum_{n=0}^N
F^{ijm}_{kln}
\Phi 
\bpm \includegraphics[scale=0.35]{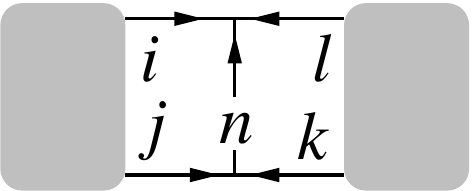} \epm
%\nonumber\\
%\Phi
%\bpm \includegraphics[scale=0.35]{Brd1O}\epm &= 
%\sum_{k=0}^N
%\om^{k}_{ij}
%\Phi
%\bpm \includegraphics[scale=0.35]{Brd2O}\epm
%\nonumber\\
%\Phi
%\bpm \includegraphics[scale=0.35]{Brd1pO}\epm &= 
%\sum_{k=0}^N
%\om^{k}_{ij}
%\Phi\bpm\includegraphics[scale=0.35]{Brd2O}\epm
\end{align}
where the shaded areas represent other parts of string-nets that are not
changed. Here, the type-$0$ string is interpreted as the no-string state. 
We would like to mention that we have drawn the first local rule
somewhat schematically. The more precise statement of this rule is that any two
string-net configurations that can be continuously deformed into each other 
have the same amplitude. In other words, the string-net wave function $\Phi$ 
only depends on the topologies of the graphs; it only depends on how 
the strings are connected (see Fig. \ref{stringnetS}).

By applying the local rules in \eqn{lclrl} multiple times, one can compute
the amplitude of any string-net configuration in terms of the amplitude of
the no-string configuration. Thus \Eq{lclrl} determines the 
string-net wave function $\Phi$. 

However, an arbitrary choice of $(d_i, F^{ijk}_{lmn})$
%, \om^{k}_{ij})$ 
does not 
lead to a well defined $\Phi$. This is because two string-net configurations 
may be related by more than one sequence of local rules. We need to choose the 
$(d_i, F^{ijk}_{lmn})$ carefully so that different sequences of 
local rules produce the same results. That is, we need to choose 
$(d_i, F^{ijk}_{lmn})$ so that the rules are self-consistent. 
Finding these special tensors is the subject of tensor category theory 
\cite{Tur94,FNS0428}. It has been shown that only those that satisfy
\cite{LWstrnet}
\begin{align}
\label{pentbrdid}
F^{ijk}_{j^*i^*0} &= \frac{v_{k}}{v_{i}v_{j}} \del_{ijk} \nonumber \\ 
F^{ijm}_{kln} = F^{lkm^*}_{jin} &= F^{jim}_{lkn^*} = F^{imj}_{k^*nl}
\frac{v_{m}v_{n}}{v_{j}v_{l}} \nonumber \\
\sum_{n=0}^N
F^{mlq}_{kpn} F^{jip}_{mns} F^{jsn}_{lkr}
&= F^{jip}_{qkr} F^{riq}_{mls}
%\nonumber\\
%\omega^{m}_{js}F^{sl^*i}_{kjm^*}\omega^{l}_{si}
%\frac{v_j v_s}{v_m} &=
%\sum_{n=0}^{N}F^{ji^*k}_{s^*nl^*}\omega^{n}_{sk}F^{jl^*n}_{ksm^*} 
%\nonumber \\
%\omega^{j}_{is} &= \sum_{k=0}^N \omega^{k}_{si^*}F^{i^*s^*k}_{isj*}
\end{align}
will result in self-consistent rules and a well defined string-net wave 
function $\Phi$. Such a wave function describes a string-net condensed state. 
Here, we have introduced some new notation: $v_i$ is defined by 
$v_i = v_{i^*} = \sqrt{d_i}$ while $\del_{ijk}$ is given by 
\begin{equation*}
\del_{ijk}=
\begin{cases}
1, & \text{if $i,j,k$ strings can join,}\\
0, & \text{otherwise}
\end{cases}
\end{equation*}
The solutions $(d_i, F^{ijk}_{lmn})$ give us a quantitative description of
topological orders (or pattern of long-range entanglements), in terms of local
dancing rules.  From the data $(d_i, F^{ijk}_{lmn})$, we can compute the
topological properties of the corresponding topological phases, such as ground
state degeneracy, quasi-particle statistics,
etc.\cite{FNS0428,LWstrnet,FFN0683,H0904,CGW1038,H1171} The above approach can
also be used to systematically describe BG topological orders in 3+1
dimensions.\cite{LWstrnet,LWuni,WW1132}

We know that group theory is the mathematical foundation of symmetry breaking
theory of phases and phase transitions.  The above systematic description of
(2+1)D BG topological order strongly suggests that tensor category theory is
the mathematical foundation of topological order and long-range entanglements.
Because of symmetry,  group theory becomes very important in physics.  Because
of quantum entanglements,  tensor category theory will becomes very
important in physics.

\section{Where to find long-range entangled quantum matter?}

In this article, we described the world of quantum phases.  We pointed out that
there are symmetry breaking quantum phases, and there are topologically ordered
quantum phases.  The topologically ordered quantum phases are a totally new
kind of phases which cannot be understood using the conventional concepts (such
as symmetry breaking, long range order, and order parameter) and conventional
mathematical frame work (such as group theory and Ginzburg-Landau theory).  The
main goal of this article is to introduce new concepts and pictures to describe
the new topologically ordered quantum phases.

In particular, we described how to use global dancing pattern to gain an
intuitive picture of topological order (which is a pattern of long range
entanglements). We further point out that we can use local dancing rules to
\emph{quantitatively} describe the  global dancing pattern (or topological
order). Such an approach leads to a systematic description of BG topological
order in terms of string-net (or spherical fusion category
theory),\cite{LWstrnet,H0904,CGW1038,GWW1017} and systematic description of 2D
chiral topological order in terms of pattern of
zeros\cite{WW0808,WW0809,R0634,SRC0706,BH0802,BH0802a,BW0932,BW1001a,LWW1024}
(which is a generalization of ``CDW'' description of FQH
states\cite{SL0604,BKW0608,SL0701,S0802,SY0802,ABK0816,ABK0875,S1002,FS1115}).

The local-dancing-rule approach also leads to concrete and explicit
Hamiltonians, that allow us to realize each string-net state and each FQH state
described by pattern of zeros.  However, those Hamiltonians usually contain
three-body or more complicated interactions, and are hard to realize in real
materials.  So here we would like to ask: can topological order be realized by
some simple Hamiltonians and real materials?

Of cause, non-trivial topological orders -- FQH states -- can be realized by 2D
electron gas under very strong magnetic fields and very low
temperatures.\cite{TSG8259,L8395} Recently, it was proposed that FQH states
might appear even at room temperatures with no magnetic field in flat-band
materials with spin-orbital coupling and spin
polarization.\cite{TMW1106,SGK1103,NSC1104,SGS1189,GNC1297} Finding such
materials and realizing FQH states at high temperatures will be an amazing
discovery. Using flat-band materials, we may even realize non-Abelian
fractional quantum Hall states\cite{MR9162,Wnab,WES8776,RMM0899} at high
temperatures.

Apart from the FQH effects, non-trivial topological order may also appear in
quantum spin systems.  In fact, the concept of topological order was first
introduced\cite{Wtop} to describe a chiral spin liquid,\cite{KL8795,WWZ8913}
which breaks time reversal and parity symmetry.  Soon after, time reversal and
parity symmetric topological order was proposed in
1991,\cite{RS9173,Wsrvb,MLB9964,MS0181} which has spin-charge separation and
emergent fermions.  The new topological spin liquid is called $Z_2$ spin liquid
or $Z_2$ topological order since the low energy effective theory is a $Z_2$
gauge theory.  In 1997, an exactly soluble model\cite{K032} (that breaks the
spin rotation symmetry) was obtained that realizes the  $Z_2$ topological
order.  Since then, the  $Z_2$ topological order become widely accepted.

More recently, extensive new numerical calculations indicated that the
Heisenberg model on Kagome lattice\cite{LBL9721,WEB9801,LE9359,JWS0803,YHW1173}
\begin{align}
 H=\sum_\text{n.n.} J \v S_i\cdot \v S_j
\end{align}
and the $J_1$-$J_2$ model on square lattice\cite{JYB1224,WGV1131,JWB1289}
\begin{align}
 H=
\sum_\text{n.n.} J_1 \v S_i\cdot \v S_j
+\sum_\text{n.n.n.} J_2 \v S_i\cdot \v S_j, \ \ \ \
J_2/J_1\sim 0.5 ,
\end{align}
may have gapped spin liquid ground states, and such spin liquids are very
likely to be $Z_2$ spin liquids.  However, with spin rotation, time reversal,
and lattice symmetry, there are many $Z_2$ spin
liquids.\cite{Wqoslpub,KLW0834,KW0906,LR1120} It is not clear which $Z_2$ spin
liquids are realized by the Heisenberg model on Kagome lattice and the
$J_1$-$J_2$ model on square lattice.  

The Heisenberg model on Kagome lattice can be realized in Herbertsmithite
$ZnCu_3(OH)_6Cl_2$.\cite{HMS0704,IFH1111} Although $J$ is as large as $150$K,
no spin ordering and other finite temperature phase transitions are found down
to 50mK.  So Herbertsmithite may realize a 2D spin liquid state.  However,
experimentally, it is not clear if the spin liquid is a gapped spin liquid or a
gapless spin liquid.  Theoretically, both a gapped $Z_2$ spin
liquid\cite{JWS0803,YHW1173,LR1120,LRL1113} and a gapless $U(1)$ spin
liquid\cite{H0013,RHL0705,HRL0813} are proposed for the Heisenberg model on
Kagome lattice.  The theoretical study suggests that the spin liquid state in
Herbertsmithite may have some very interesting characteristic properties: A
magnetic field in $z$-direction may induce a spin order in
$xy$-plane,\cite{RKL0774} and an electron (or hole) doping may induce a charge
$4e$ topological superconductor.\cite{KLW0902} 

To summarize, topological order and long-range entanglements give rise to new
states of quantum matter.  Topological order has many new emergent phenomena,
such as emergent gauge theory, fractional charge, fractional statistics,
non-Abelian statistics, perfect conducting boundary, etc.  In particular, if we
can realize a quantum liquid of oriented strings in certain materials, it will
allow us to make artificial elementary particles (such as artificial photons
and artificial electrons).  So we can actually create an artificial vacuum, and
an artificial world for that matter, by making an oriented string-net liquid.
This would be a fun experiment to do!

\section{A new chapter in physics}

Our world is rich and complex. When we discover the inner working of our world
and try to describe it, we ofter find that we need to invent new mathematical
language describe our understanding and insight.  For example, when Newton
discovered his law of mechanics, the proper mathematical language was not
invented yet.  Newton (and Leibniz) had to develop calculus in order to
formulate the law of mechanics.  For a long time, we tried to use theory of
mechanics and calculus to understand everything in our world.

As another example, when Einstein discovered the general equivalence principle
to describe gravity, he needed a  mathematical language to describe his theory.
In this case, the needed mathematics, Riemann geometry, had been developed,
which leaded to the theory of general relativity.  Following the idea of
general relativity, we developed the gauge theory. Both general relativity and
gauge theory can be described by the mathematics of fiber bundles.  Those
advances led to a beautiful geometric understanding of our would based quantum
field theory, and we tried to  understand everything in our world in term of
quantum field theory.

Now, I feel that we are at another turning point.  In a study of quantum
matter, we find that long-range entanglements can give rise to many new quantum
phases.  So long-range entanglements are natural phenomena that can happen in
our world.  But  mathematical language should we use to describe long-range
entanglements?  The answer is not totally clear. But early studies suggest that
tensor category and group cohomology should be a part of the mathematical frame
work that describes long-range entanglements.  What is surprising is that such
a study of quantum matter might lead to a whole new point of view of our world,
since long-range entanglements can give rise to both gauge interactions and
Fermi statistics. (In contrast, the geometric point of view can only lead to
gauge interactions.) So maybe we should not use geometric pictures, based on
fields and fiber bundles, to understand our world.  Maybe we should use
entanglement pictures to understand our world.  We may live in a truly quantum
world.  So, quantum entanglements may represent a new chapter in physics.

\bibliography{../../bib/wencross,../../bib/all,../../bib/publst} 
\end{document}